\documentclass[aps,pre,twocolumn,superscriptaddress,longbibliography]{revtex4-2}

\usepackage{graphicx, subfigure}
\usepackage{amssymb, amsmath,amssymb,amsfonts}
\usepackage{amsthm,mathrsfs,amsopn}
\usepackage{dcolumn}
\usepackage{bm}
\usepackage{color}
\usepackage[utf8]{inputenc}
\usepackage{mathtools}
\usepackage[normalem]{ulem}

\definecolor{green}{rgb}{0.,0.5,0.5}

\definecolor{orange}{rgb}{0.,0.5,0.5}

\def\i{\textrm{i}}

\newcommand{\be}{\begin{equation}}
\newcommand{\ee}{\end{equation}}
\newcommand{\bea}{\begin{eqnarray}}
\newcommand{\eea}{\end{eqnarray}}

\def\bc{\begin{center}}
\def\ec{\end{center}}

\newcommand{\Avg}[1]{\left\langle{#1}\right\rangle}

\theoremstyle{plain} 

\definecolor{brickred}{rgb}{0.7, 0.25, 0.33}
\definecolor{applegreen}{rgb}{0.55, 0.71, 0.0}

\begin{document}

\title{Cup and cap  synchronization: higher-order node dynamics from topology}
\author{Runyue Wang}
\affiliation{School of Mathematical Sciences, Queen Mary University of London, London, E1 4NS, United Kingdom}
\author{Yuanzhao Zhang} 
\affiliation{Department of Physics and Astronomy, University of Rochester, Rochester, New York, USA}
\author{Ginestra Bianconi}
\affiliation{School of Mathematical Sciences, Queen Mary University of London, London, E1 4NS, United Kingdom}

\begin{abstract}
Higher-order node-based synchronization provides distinct mechanisms for discontinuous synchronization transitions. However, a comprehensive theoretical framework capable of encompassing existing models and revealing the role of higher-order interactions beyond triangles is still missing. Here, combining algebraic topology and nonlinear dynamics, we introduce Cup and Cap Synchronization, a unifying theoretical framework for higher-order node-based dynamics that naturally incorporates the topological organization of the underlying higher-order network. We show that Cup and Cap Synchronization can display critical behavior belonging to the universality classes of two existing higher-order Kuramoto-like models, whose distinct critical behaviors were thought to originate from different mechanisms. Remarkably, we demonstrate that Cup and Cap Synchronization displays additional phenomenology not captured by the previously proposed models of higher-order node-based synchronization. Most relevantly, the inclusion of four-body interactions can fundamentally alter the critical properties of the synchronization transition, giving rise to collective phenomena that cannot be observed in models including only interactions mediated by edges and triangles. These findings reveal the importance of going beyond the commonly studied three-body terms and establish Cup and Cap Synchronization as a unifying theoretical framework for understanding collective dynamics in higher-order networks.
\end{abstract}

\maketitle
Synchronization~\cite{kuramoto1975,arenas2008synchronization,rodrigues2016kuramoto,strogatz2000kuramoto,acebron2005kuramoto,schaub2016graph,witthaut2017classical,restrepo2005onset,montanari2026disorder,chandra2019continuous}, as  captured by the Kuramoto model, describes the onset of dynamical order, a  collective behavior of fundamental importance for complex systems from the brain to biological rhythms.
The Kuramoto model was traditionally formulated for synchronization on networks with pairwise interactions.
However, complex systems include nonpairwise interactions~\cite{battiston2021physics}, making higher-order networks~\cite{bianconi2021higher,bick2023higher,salnikov2019simplicial,torres2021and,rosas2022disentangling,majhi2022dynamics}, including simplicial complexes and hypergraphs, their natural representation.
Higher-order networks have been shown to display rich collective behavior for diverse dynamical processes, including random walks~\cite{carletti2020random},  diffusion ~\cite{torres2020simplicial}, consensus dynamics~\cite{neuhauser2020multibody}, epidemic spreading~\cite{iacopini2019simplicial}, and synchronization~\cite{tanaka2011multistable,skardal2019abrupt,bick2016chaos,skardal2020higher,zhang2021unified,gallo2022synchronization,gambuzza2021stability,kuehn2021universal,kundu2022higher}.

Specifically, higher-order node-based synchronization \cite{tanaka2011multistable,skardal2019abrupt,bick2016chaos,skardal2020higher,lucas2020multiorder} has attracted considerable interest~\cite{battiston2026collective}. While significant attention has been devoted to investigating model stability~\cite{dai2025higher} and the role of structural nestedness in determining the properties of higher-order synchronization \cite{zhang2023higher,muolo2026synchronization,lamata2025hyperedge}, important theoretical questions remain open.

Kuramoto-like, higher-order node-based synchronization can be broadly classified into two different classes of models, both displaying a discontinuous transition. The first model,  proposed in Refs.~\cite{tanaka2011multistable,skardal2019abrupt}, displays a strong dependence on the bias externally imposed by initial conditions, with only the desynchronization transition displaying a discontinuity. In contrast, the second model, proposed in \cite{bick2016chaos,skardal2020higher}, displays a stable hysteresis loop.
Both the origin of these two models and their physical motivation, beyond the  different mechanisms proposed, have been the subject of  intense debate \cite{battiston2026collective,bick2024higher,leon19,leon2024higher}.
However, so far, whether the  distinct critical  behaviors of these  two models  might be understood in terms of a unifying theoretical framework is a question that remains unanswered.

The key property of higher-order networks is their power to represent the complex system in terms of its complete higher-order topological structure. Specifically, topology  reveals properties of data not  obtainable with combinatorial and statistical measures relying only on pairwise data. For instance, topological data analysis can distinguish between different functional states of a complex system~\cite{petri2014homological,petri2013topological}.

Topology is however not only relevant to characterize the structure of complex systems but also to describe their dynamics. Indeed, topology has been shown to shape the dynamics of higher-order topological signals, i.e. dynamical variables associated with edges, triangles, and higher-order simplices~\cite{millan2020explosive}.  The Topological Synchronization model \cite{millan2020explosive,ghorbanchian2021higher,arnaudon2022connecting,bavcic2026phase,zaid2025designing,guo2026topological}, together with Global Topological Synchronization \cite{carletti2023global,wang2024global,wang2026topology},  combines algebraic topology with nonlinear dynamics revealing not only that topology shapes higher-order dynamics, but also that dynamics can learn topology \cite{millan2025topology}. Thus, a crucial question is: can topology elucidate as well the higher-order mechanisms for higher-order node-based dynamics?

Here, we propose the Cup and Cap Synchronization model as a unified theoretical framework for higher-order node-based dynamics based on topology. This model adopts the cup and cap products, which have already been proven to be relevant in Topological Deep Learning~\cite{niedostatek2026cup}, as natural algebraic-topological operators for describing node-based interactions mediated by higher-order simplices. The Cup and Cap Synchronization model describes node-based dynamics mediated by higher-order interactions of arbitrary dimension, here studied up to $(M+1)$-body interactions with $M=2,3$ (i.e., up to triangles and tetrahedra, respectively). For $M=2$  this model provides a unified theoretical framework for higher-order node-based synchronization, encompassing previously proposed models~\cite{tanaka2011multistable,skardal2019abrupt,bick2016chaos,skardal2020higher} as limiting cases, while extending to a broad class of dynamical processes whose critical behavior belongs to either of their universality classes. Thus, the Cup and Cap Synchronization model reveals that  the same theoretical framework can account for two dynamical and critical behaviors of higher-order synchronization, previously thought to be originating from distinct mechanisms~\cite{tanaka2011multistable,skardal2019abrupt,bick2016chaos,skardal2020higher}.

Interestingly, Cup and Cap Synchronization reveals features of higher-order node-based synchronization that go beyond those captured by previously proposed models. In certain regions of parameter space, the dynamics depends on half-integer order parameters, exhibiting a $4\pi$-periodicity. Although the dynamics considered here is entirely classical, $4\pi$-periodicity also arises in quantum systems, notably in topological Josephson junctions and fractional spin Josephson effects~\cite{fu2009josephson,pientka2013magneto}. Importantly, the $4\pi$-periodicity associated with the half-integer order parameters is not merely a rescaling of the phases, since their definition requires a choice of branch cut that distinguishes phases differing by $2\pi$, introducing an additional structure into the synchronization dynamics. As a consequence, Cup and Cap Synchronization can exhibit a strong dependence on the initial phase distribution, giving rise to desynchronization dynamics characterized by imaginary half-integer order parameters. This behavior reveals a sensitivity to the initial phase bias that has no counterpart in previously proposed models~\cite{tanaka2011multistable,skardal2019abrupt}.
Moreover, and most importantly, for $M=3$ we show that the Cup and Cap Synchronization model 
can display a critical behavior in a distinct universality class that cannot be observed in models including only pairwise and three-body interactions.
In this scenario, the model simultaneously displays a stable hysteresis loop and a strong dependence on the initial conditions along the desynchronization dynamics.

\section{Results}
\subsection{Background on the cap and cup products}
The cup and cap products are ideal algebraic topological tools able to lift node signals to arbitrary order~\cite{niedostatek2026cup}, thus encompassing limitations of approaches based on the boundary operator that can only be used to lift node signals to the edges.
Due to this remarkable ability to lift node signals, the cap and cup products are here adopted to formulate the unifying Cup and Cap Synchronization model of higher-order node-based synchronization that we propose here. 
\begin{figure}[!t]
    \includegraphics[width=0.98\columnwidth]{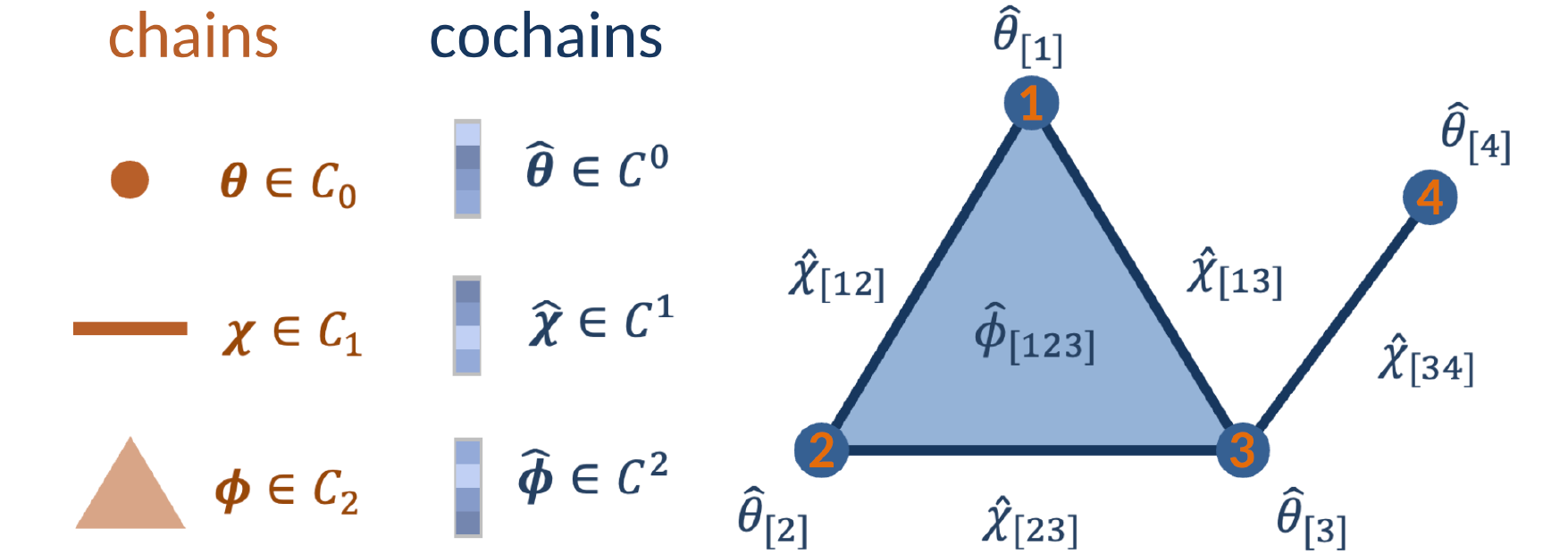}
    \caption{\textbf{Chains and cochains  encode the higher-order topology and  dynamics of simplicial complexes.} A $0$-chain $\boldsymbol{\theta} \in C_0$ is a linear combination of nodes. The corresponding $0$-cochain $\hat{\boldsymbol{\theta}} \in C^0$ is a linear combination of functions that are defined on nodes. The same construction on the links and triangles gives a $1$-chain ($\boldsymbol{\chi} \in C_1$) and a $2$-chain ($\boldsymbol{\phi} \in C_2$) with associated $1$-cochain ($\hat{\boldsymbol{\chi}} \in C^1$) and $2$-cochain ($\hat{\boldsymbol{\phi}} \in C^2$).}
    \label{fig1}
\end{figure}
We consider a $d$-dimensional simplicial complex $\mathcal{K}$ defined as a set of simplices among $N$ nodes. We indicate with $\mathcal{Q}_p$ the set of $p$-simplices in $\mathcal{K}$. Specifically, we  consider simplices with two different positive orientations, and we indicate with ${\bf e}_{[i_0 i_1 i_2 \ldots i_p]}$ their canonical basis. In algebraic topology, $p$-chains $C_p$ are linear combinations of $p$-simplices.
The $p$-cochains $C^p$ are functions defined on $p$-simplices, or more specifically on $p$-chains, thus cochains are the natural objects to study the dynamics on higher-order networks using an algebraic topological language (see Fig.~\ref{fig1} for an illustration of chains and cochains).
The $p$-cochains defined on top of simplices  admit as canonical basis the canonical functions defined on these simplices here indicated as $\hat{\bf e}_{[i_0 i_1 i_2 \ldots i_p]}$.  
We indicate chains with  Greek letters such as $\bm\theta$,  and their associated cochains with a hat symbol, i.e. $\hat{\bm\theta}$.
For instance, if $\bm\theta$ indicates a $0$-chain and $\hat{\bm\theta}$ its associated $0$-cochain we have 
\bea
\bm\theta=\sum_{i=1}^N\theta_{[i]}{\bf e}_{[i]},\quad \hat{\bm\theta}=\sum_{i=1}^N\theta_{[i]} \hat{\bf e}_{[i]}
\eea
A similar convention is adopted for higher-dimensional chains and cochains.

\subsubsection*{Cup product as lifting operator}
The cup product can be used to lift pairs of cochains to a cochain whose dimension is the sum of the dimensions of the cochains from which it is built.
Consider the $p$-cochain $\hat{\bm\theta}\in C^p$ and the $q$-cochain $\hat{\bm\chi}\in C^q$. The cup product $\hat{\bm\phi}=\hat{\bm\theta}\smile \hat{\bm\chi}\in C^{p+q}$ is the $(p+q)$-cochain whose elements are 
\bea
\hspace{-6mm}\hat{\bm\theta}\smile\hat{\bm\chi}&=&\hat{\bm\phi}=\sum_{[i_0 i_1 \ldots i_{p+q}]\in \mathcal{Q}_{p+q}}\phi_{[i_0 i_1 \ldots i_{p+q}]}\hat{\bf e}_{[i_0 i_1 \ldots i_{p+q}]}
\eea
with 
\bea
\phi_{[i_0 i_1 \ldots i_{p+q}]}=\theta_{[i_0 i_1 \ldots i_p]}\chi_{[i_p i_{p+1}\ldots i_{p+q}]}
\eea
Thus, the cup product can be used to lift the cochains to higher-order.

\subsubsection*{Cap product as lowering operator}
The cap product can be used to transform a given cochain to a chain of lower dimension.
Consider a $q$-chain ${\bm{\psi}}\in C_q$ and a $p$-cochain $\hat{\bm{\theta}}\in C^p$ with $q\geq p$. 
The cap product $\hat{\bm{\theta}}\frown {\bm{\psi}}\in C_{q-p}$ is the $q-p$ chain of elements
\bea
\hspace*{-5mm} \hat{\bm{\theta}}\frown {\bm{\psi}}=\sum_{[i_0 i_1 \ldots i_q]\in \mathcal{Q}_q}\theta_{[i_0 i_1 \ldots  i_p]}\psi_{[i_0 i_1 \ldots i_q]}{\bf e}_{[i_p i_{p+1} \ldots  i_{q}]}
\eea
Thus,  the cap product can be used to lower the cochains leading to lower dimensional chains.

For more details on the cup and cap products, see Refs.~\cite{nanda2021computational,hatcher2005algebraic,munkres2018elements}.

\subsection{Cup and cap product formulation of the standard Kuramoto model}
The cup and cap products provide a natural language to formulate the standard Kuramoto model.
Let us illustrate this finding by considering the Kuramoto model defined on a  network $G$. This model is well known to describe the synchronization of phases $\theta_{[i]}$ associated with each node $i$ of the network, when each node is assigned an intrinsic frequency $\omega_{[i]}$ drawn from a unimodal distribution $g(\omega)$, typically taken to be either  a Gaussian or  Lorentzian distribution.
Indicating with  $a_{[ij]}$ (with $a_{[ij]}=a_{[ji]}$)  the  adjacency matrix entries of the network $G$,  and with $\kappa>0$ the coupling constant, the dynamical equations of the Kuramoto model read,
\bea
   \frac{d\theta_{[i]}}{dt} = \omega_{[i]}+\kappa\sum^N_j a_{[ij]}\sin{(\theta_{[j]}-\theta_{[i]})}.
   \label{stdKuramoto}
   \eea
   Here we want to show that this equation admits an equivalent algebraic topology formulation expressed in terms of the cup and cap products.
  We consider the $0$-cochains $\hat{\bm\theta},\hat{\bm\omega}\in C^0$ encoding   the  phases $\theta_{[i]}$,  and the intrinsic frequencies $\omega_{[i]}$ associated with each node $i$ of the simplicial complex and their associated chains $\bm\theta$ and $\bm\omega$  defined as 
  \bea
  \bm\theta=\sum_{i=1}^N\theta_{[i]}{\bf e}_{[i]},\quad \hat{\bm\theta}=\sum_{i=1}^N\theta_{[i]} \hat{\bf e}_{[i]}\nonumber \\
\bm\omega=\sum_{i=1}^N\omega_{[i]}{\bf e}_{[i]},\quad \hat{\bm\omega}=\sum_{i=1}^N\omega_{[i]} \hat{\bf e}_{[i]}
  \eea
Starting from the $0$-cochain $\hat{\bm\theta}$, encoding the phases associated with the nodes, we define the 1-cochain $\hat{\bm\eta}\in C^1$ as the 1-cochain $\delta \hat{\bm\theta}$ where $\delta:C^0\to C^1$ is the coboundary operator given by 
\bea
\hat{\bm\eta}=\delta \hat{\bm\theta}=\sum_{i=1}^N\sum_{j=1}^N\eta_{[ij]}\hat{\bf e}_{[ij]},
\label{eq:eta1}
\eea
with 
\bea
\eta_{[ij]}=\theta_{[j]}-\theta_{[i]}.
\label{eq:eta2}
\eea
Applying the $\sin(\hat{\bm\eta})$ function element-wise we obtain the $1$-cochain
\bea
\sin(\hat{\bm\eta})=\sum_{i=1}^N\sum_{j=1}^N\sin(\theta_{[j]}-\theta_{[i]})\hat{\bf e}_{[ij]}
\eea
We also construct the $1$-chain $\bm\alpha$ built from the adjacency matrix elements of the network defined as \bea
\bm\alpha^{(1)}=\sum_{i=1}^N\sum_{j=1}^Na_{ij}{\bf e}_{ij}.\eea
With this notation the standard Kuramoto model (\ref{stdKuramoto}) can  equivalently be written in terms of the cup and cap product as
\bea
\frac{d{\bm\theta}}{dt}={\bm\omega}+\kappa \sin(\hat{\bm\eta})\frown{\bm \alpha}^{(1)}.
\label{eq:top_st_K}
\eea
This expression indicates the lifting of the node signal via the coboundary, leading to the $1$-cochain $\sin(\hat{\bm\eta})$ and the subsequent lowering of the signal with the cap product with the adjacency $1$-chain $\alpha$.
We note that, since the standard Kuramoto model only admits lifting and lowering of the node signal up to edges, this model admits two equivalent topological expressions: one in terms of the cup and cap product, as discussed above, and one in terms of the boundary and coboundary operators~\cite{millan2020explosive}.
Topological Synchronization represents the general framework that builds on the Kuramoto model expression in terms of the boundary and coboundary operators to define synchronization of  higher-order topological signals~\cite{millan2020explosive,carletti2023global,millan2025topology}.
Here, we demonstrate the complementary result, that Cup and Cap Synchronization,   building on the  expression of the standard Kuramoto model in terms of the cup and cap products,  provides a powerful way to formulate a unified topological theory of node-based higher-order synchronization.
\begin{figure}
\centering
    \includegraphics[width=0.95\columnwidth]{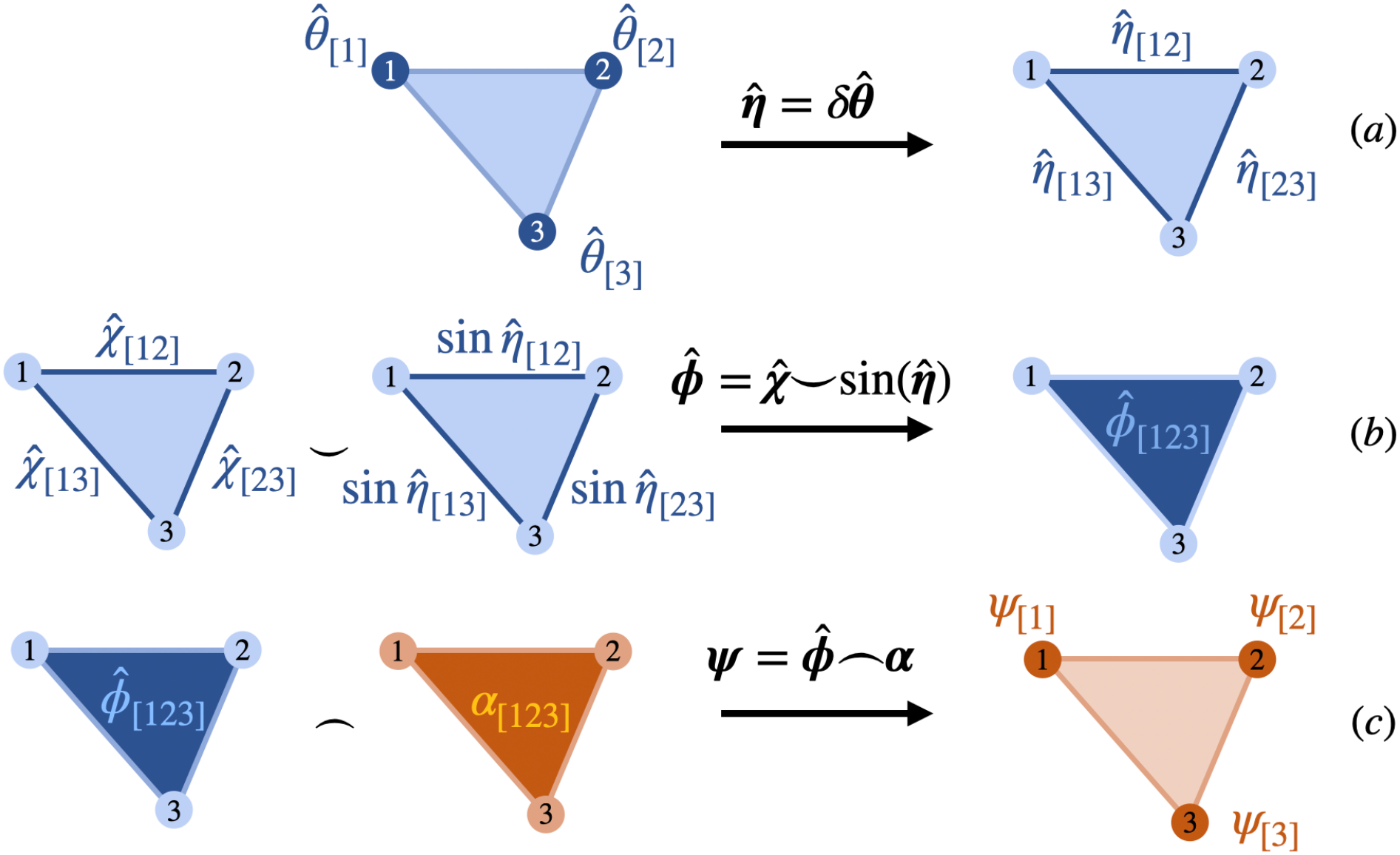}
    \caption{\textbf{Illustration  of the cup and cap products as lifting and lowering operators that allow the formulation of the Cup and Cap Synchronization model.}  Panel (a) shows that the coboundary operator (gradient) allows the node signal to be lifted to edges, as it maps a $0$-cochain to a $1$-cochain. Panel (b) shows that the cup product is a lifting operator that allows  two $1$-cochains, eventually built starting from a node signal, to be lifted up to triangles. Indeed the cup product allows the  construction of a $2$-cochain from two $1$-cochains. Panel (c) shows that the cap product is a lowering operator that can be used to lower the signal lifted to triangles back to the nodes. Specifically, the cap operator can be used to  construct a $0$-chain from a $2$-cochain and a $2$-chain.}
    \label{fig:2}
\end{figure}

\subsection{Cup and cap synchronization of node-based dynamics}
The Cup and Cap Synchronization is a  theoretical framework that unifies node-based synchronization using an algebraic topology formulation.
As we will see, this model has the ability to reconcile and include collective phenomena previously thought to be associated with distinct mechanisms, demonstrating that they are actually emerging from a single comprehensive theoretical framework combining topology with dynamics. 
Moreover, the Cup and Cap Synchronization shows distinct critical behavior when 4-body interactions are also included, which cannot be observed if only pairwise and 3-body interactions are considered.

The Cup and Cap Synchronization dynamics is a general theoretical framework for node-based dynamics that takes into account the higher-order interactions in the simplicial complex. In particular, the cup and cap products are used to lift the $0$-cochain under consideration to arbitrary order and then to lower the dynamics back to zero order (dynamics defined on the nodes). In this way, the resulting dynamics takes full advantage of the simplicial complex structure of the interactions present in the complex system under consideration. The synchronization is achieved for coupling constant values above the synchronization threshold, which determines the onset of the collective behavior.

In order to define the Cup and Cap Synchronization on a $d$-dimensional simplicial complex $\mathcal{K}$, we encode its structure into the chains ${\bm \alpha}^{(n)}\in C_n$
 with $0\leq n\leq d$, i.e.
 \bea
 {\bm\alpha}^{(n)}=\sum_{[i_0 i_1 i_2 \ldots  i_n]\in\mathcal{Q}_{n}} a_{[i_0 i_1 i_2 \ldots  i_n]}{\bf e}_{[i_0 i_1 i_2\ldots i_n]}
 \eea
 where $a_{[i_0 i_1 i_2 \ldots  i_n]}$ indicates the entries of the adjacency tensors, i.e. $a_{[i_0 i_1 i_2 \ldots i_n]}=1$ if the corresponding simplex belongs to the simplicial complex, i.e. ${[i_0 i_1 i_2 \ldots i_n]}\in \mathcal{Q}_{[n]}$ and  $a_{[i_0 i_1 i_2 \ldots i_n]}=0$ otherwise.
 Therefore, $a_{[i_0 i_1 i_2 \ldots  i_n]}$ is invariant under any permutation $\pi(x)$ of the indices (i.e. $a_{[i_0 i_1 i_2 \ldots i_n]}=a_{\pi([i_0 i_1 i_2\ldots i_n])}$) .

The Cup and Cap Synchronization dynamics lifts the zero cochain $\hat{\bm\theta}\in C^0$ indicating the phases of the oscillators, up to  order $M$, with $M\leq d$ and then lowers it back to the nodes (see Fig.~\ref{fig:2} for an illustration of the case $M=2$).
Indicating with $\hat{\bm\omega}$ the vector of the intrinsic frequencies of the nodes each drawn from the $g(\omega)$ distribution, and  with $\bm\theta\in C_0$ the chain associated with $\hat{\bm\theta}\in C^0$, the Cup and Cap Synchronization dynamics reads, in full generality,
\bea
\frac{d{\bm\theta}}{dt}={\bm\omega}+\sum_{n=1}^M {\kappa^{(n)}} (\hat{\bm\chi}^{(n-1)}\smile\sin(\hat{\bm\eta}))\frown{\bm \alpha}^{(n)},
\label{eq:ccs}
\eea
where $\kappa^{(n)}$ are the order-dependent coupling constants.
Here we take $\hat{\bm\chi}^{(n-1)}\in C^{n-1}$ to be auxiliary $(n-1)$-cochains built starting from $\hat{\theta}$,  specifically chosen to be functions of $\hat{\bm\eta}$ defined in Eqs.(\ref{eq:eta1})-(\ref{eq:eta2}), i.e. we have for $n>1$
\bea
\hat{\bm\chi}^{(n-1)}=f_{n-1}(\hat{\bm\eta}/2)\smile f_n(\hat{\bm\eta}/2)\smile \ldots \smile f_1(\hat{\bm\eta}/2)
\eea
where $f_n(\hat{\bm\eta}/2)$ are  even functions taking element-wise values,  while for $n=1$ we put  $\hat{\bm\chi}^{(0)}={\bf 1}$ by definition.
We focus specifically on the choice  $f_n(x)=[\cos(x)]^{h_n-1}$  with $h_n$ being non-negative integers, i.e. $h_n\geq 0$.
Note that the Cup and Cap Synchronization reduces to the standard Kuramoto model for simplicial complexes for $M=1$, and that its dynamical Eqs.~$(\ref{eq:ccs})$ are invariant under the gauge transformation 
\bea
\bm\theta \to \bm\theta-\Omega t{\bf 1},\quad  \bm\omega \to  \bm\omega-\Omega{ \bf 1}, 
\eea
as it is natural to require. Thus, we can always choose the usual gauge  $\Avg{\omega_{[i]}}=0$.
\subsection{Cup and cap Synchronization mediated by edges and triangles}
We first focus on the relevant case of a fully connected simplicial complex with $M=2$, i.e. including interactions mediated by nodes, edges and triangles. In this setting, the Cup and Cap Synchronization dynamics can be written element-wise as 
\bea
   \frac{d\theta_{[i]}}{dt} &=& \omega_{[i]}+\frac{\sigma_1}{N}\sum_{j=1}^N a_{[ij]} \sin{(\theta_{[j]}-\theta_{[i]})}\nonumber \\&&\hspace*{-7mm}+\frac{\sigma_2}{N^2}\sum_{j=1}^N\sum_{k=1}^Na_{[ijk]}\sin(\theta_{[j]}-\theta_{[i]})f(\eta_{[kj]}/2),
   \label{cupKuramoto}
   \eea
   where $\sigma^{(n)}$ are the rescaled coupling constants.
   As mentioned before, we take  $f(x)=f_1(x)=[\cos(x)]^{h_1-1}$  with $h_1=h$ being a non-negative integer, i.e. $h\geq 0$. In this setting  the Cup and Cap Synchronization dynamics becomes
\bea 
    \frac{d\theta_{[i]}}{dt} &=& \omega_{[i]}+\frac{\sigma_1}{N}\sum_{j=1}^N a_{[ij]} \sin{(\theta_{[j]}-\theta_{[i]})}\nonumber \\&&\hspace{-5mm}+\frac{\sigma_2}{N^2}\sum_{j=1}^N\sum_{k=1}^Na_{[ijk]} W_{[ijk]}(\theta_{[i]},\theta_{[j]},\theta_{[k]}),
    \label{dynamics_phase_M2}
\eea
where the higher-order coupling term $W_{[ijk]}(\theta_{[i]},\theta_{[j]},\theta_{[k]})$ reads
\bea
\hspace{-4mm}W_{[ijk]}(\theta_{[i]},\theta_{[j]},\theta_{[k]})=\sin{\left( \frac{\theta_{[k]} + \theta_{[j]}}{2} - \theta_{[i]}\right)} \left( \cos{\frac{\theta_{[k]}-\theta_{[j]}}{2}}\right)^{h}\nonumber \\\hspace{-10mm}=\frac{1}{2^{h}}\sum_{n=0}^h \binom{h}{n} \sin \left(\frac{2n-h+ 1}{2} \theta_{[k]} - \frac{2n-h-1}{2} \theta_{[j]} - \theta_{[i]} \right)\nonumber
\eea
From this expression it is clear that the dynamical equations have a period $2\pi$ for $h$ odd. For even $h$, however, the higher-order interactions naturally induce a $4\pi$-periodic phase dynamics, distinguishing configurations that differ by a $2\pi$ phase shift. Importantly, as we will show below, the $4\pi$-periodicity of the half-integer order parameter cannot be reduced to a simple rescaling of the phases, since it involves a choice of branch cut that distinguishes phases separated by $2\pi$, revealing an additional structure underlying the synchronization dynamics.  This extended periodicity is a distinctive feature of the model and gives rise to half-integer order parameters that capture additional information about the collective dynamics.

The Cup and Cap Synchronization model provides a unified theoretical framework for higher-order node-based synchronization, encompassing previously proposed models as limiting cases (see Methods). Specifically, for $h=0$ and $\sigma_1=0$, a rescaling of the phases recovers the model of Refs.~\cite{tanaka2011multistable,skardal2019abrupt}, while for $h=3$, the dynamics can be recast into the model of Refs.~\cite{bick2016chaos,skardal2020higher}. Beyond these limiting cases, the framework describes a broader class of higher-order synchronization dynamics, extending the range of possible collective behaviors and critical phenomena.

In the following, we develop an analytical theory of the Cup and Cap Synchronization model, which will lead to the characterization of  its collective dynamics and critical properties across the different regimes.
Neglecting subleading terms in $1/N$, the dynamical equations (\ref{dynamics_phase_M2}) can be  expressed in terms of the complex parameters $X_{p/2}$, defined as
\bea
 \hat{X}_{p/2} ={R}_{p/2}e^{\textrm{i}\Psi_{p/2}}=\frac{1}{N} \sum_{j=1}^N e^{\textrm{i}p \theta_{[j]}/2}.
\eea
In particular, the Cup and Cap Synchronization defined by Eqs.(\ref{dynamics_phase_M2}) reduce to 
\bea
\dot{\theta}_{[i]} = \omega_{[i]} - \text{Im}\left[{\Lambda} \exp(\textrm{i}\theta_{[i]})\right]
\label{eq:dyn_Lambda}
\eea
where $\Lambda$ is given by
\bea
{\Lambda} = \left(\sigma_1 \Theta^{(1)} + \frac{\sigma_2}{2^h} \Theta^{(2)}\right)^*\eea
with $^*$ indicating complex conjugation and  $\Theta^{(1)}=X_1$, and $\Theta^{(2)}$ given by
\bea
\Theta^{(2)}=\sum_{n=0}^h \binom{h}{n} {X}_{(2n-h+1)/2} {X}_{-(2n-h-1)/2}.
\label{eq:Theta2}
\eea
For $h$ odd $\Theta^{(2)}$ only includes products $X_q$ with $q\in \mathbb{Z}$ (which we indicate for short as $X_{\mathbb{Z}}$). For $h$ even, instead, it includes a pair of half-integer order parameters $X_{p/2}$ with $p=2q+1$ with $q\in\mathbb{Z}$ which we indicate as $X_{\mathbb{Z}/2}$.
We note that due to the gauge symmetry of the dynamics, $\Lambda$ can always be chosen to be real.
The continuity equation for the probability density function $\rho(\theta,t|\omega)$ can be derived from the dynamics described by Eq.~(\ref{eq:dyn_Lambda}) as 
\begin{equation}
    \partial_t \rho + \partial_\theta (\rho v) = 0,
\end{equation}
with velocity 
\begin{equation}
    v(\theta) = \omega +\textrm{i} \frac{{\Lambda}}{2} \left( e^{\textrm{i} {\theta}} -  e^{-\textrm{i} {\theta} }\right).
\end{equation}
\begin{figure*}
    \includegraphics[width=1\textwidth]{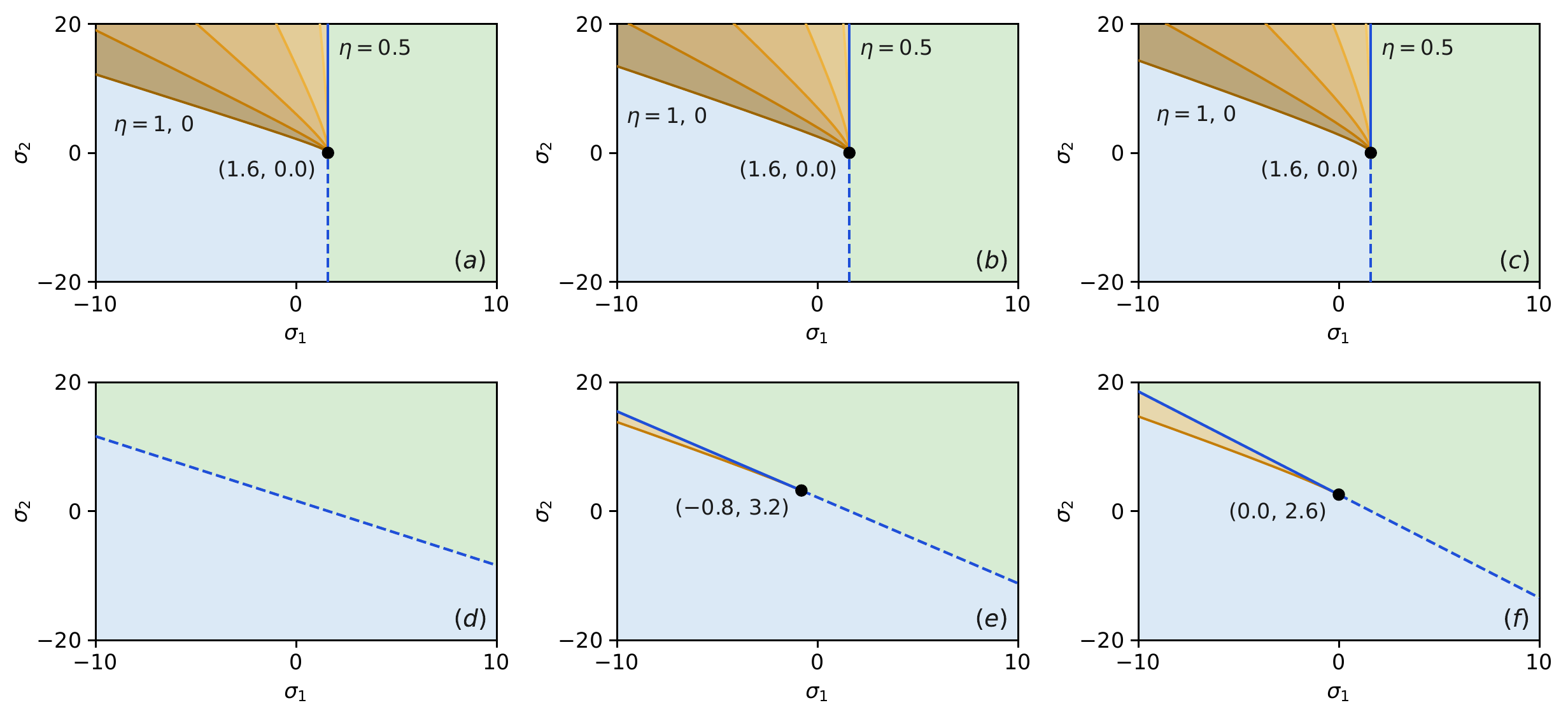}
  \caption{\textbf{The phase diagrams of the Cup and Cap Synchronization model for $M=2$ reveal a distinct critical behavior for $h$ even and $h$ odd.} For $h$ even (panels (a)-(c)) the synchronization transition remains continuous, and  the desynchronization transition has a remarkable dependence on the bias $\eta$. For $h$ odd  (panels (d)-(f)) and as long as $h\neq 1$, the model displays a saddle-node and a pitchfork bifurcation leading to a region of its phase diagram with a stable hysteresis loop. The diagrams indicate synchronized (green), bistable (brown), and incoherent (blue) states, as a function of coupling constants $\sigma_1$ and $\sigma_2$, on a fully connected network with Gaussian-distributed intrinsic frequency $\omega$.   Panels (a)-(c) correspond to $h=0,2,4$. Brown lines, from dark to light, correspond to the saddle-node bifurcations with bias $\eta,(1-\eta)=1,0.9,0.8,0.7,0.6,0.5$. Panels (d)-(f) correspond to $h=1,3,5$. These phase diagrams correspond to  a dynamics defined on a fully connected network with a  Gaussian intrinsic frequency distribution. For $h$ even, the bias $\eta$ indicates the probability that a given phase satisfies $\theta\in (-\pi,\pi]$ while $1-\eta$ indicates the probability that $\theta\in (\pi,3\pi]$.}
    \label{fig:phasediagram}
\end{figure*}
Since the dynamical Eqs. (\ref{dynamics_phase_M2}) are periodic under the transformation $\theta_{[i]}\to \theta_{[i]}+4\pi$ for any generic oscillator $i$,  we consider the following Fourier expansion for $\rho(\theta, t|\omega)$
\bea
    \rho(\theta, t|\omega) = \frac{1}{4\pi}\left(1 + \sum_{m=1}^\infty \hat{f}^{(m)}(\omega,t) e^{\textrm{i}m\theta/2} + c.c.\right)
\eea
where we have $\rho(\theta,t|\omega) = \rho(\theta + 4\pi,t|\omega)$.
The Ott-Antonsen ansatz holds for the symmetric part of the density distribution $\rho(\theta, t|\omega)$, i.e.
\begin{equation}
    \hat{f}^{(2m)}(\omega, t)  = [{c}(\omega,t)]^{2m},
\end{equation}
while the equivalent assumption  for the antisymmetric part does not hold in this case. This necessary distinction corresponds, with the necessary modifications, to the similar distinction between the even and odd behavior of the probability density function required to solve the model proposed in Refs.~\cite{tanaka2011multistable,skardal2019abrupt}. 
At stationarity, we thus find 
\bea
\Big(c(\omega)\Big)^2 = -\mathrm{i}\frac{\omega}{{\Lambda}} \pm \sqrt{1 - \left(\frac{\omega}{{\Lambda}}\right)^2}
\eea
It follows that, when $k$ is a positive integer, we have
\bea
R_k=R_{-k}=\int d\omega g(\omega) \left(c_r^*(\omega)\right)^{2k}
\eea
By contrast, the half-integer parameters  $X_{p/2}$ with $p$ odd require the explicit calculation of the integral, 
\bea
    X_{p/2}=\int \text{d}\omega g(\omega)\int_0^{4\pi} \text{d}\theta \rho(\theta, t|\omega) e^{\textrm{i}p\theta/2} 
\eea
This integral can be evaluated using  the even part $\rho(\theta, t| \omega)$ only if we know the fraction of nodes $\eta$ with phases $\theta\in (-\pi,\pi]$. Therefore, this observation reveals that the dynamics can be modified by biasing the initial condition of the model, which is particularly relevant for the desynchronization dynamics.
As an illustration,  we provide here the explicit expression for $X_{1/2}$
\bea
X_{1/2}&=&(2\eta-1)\int_{|\omega|\leq \Lambda} d\omega g(\omega) \sqrt{\frac{1}{2}\left[1+\sqrt{1-\left(\frac{\omega}{\Lambda}\right)^2}\right]}\nonumber
\eea
where $\eta$ indicates the fraction of nodes with phases in $(-\pi,\pi]$ and $1-\eta$ the fraction of nodes with phases in $(\pi,3\pi]$ (for the explicit expression of all the order parameters used in this study, see the SI).

\begin{figure}[htbp]
\includegraphics[width=1\columnwidth]{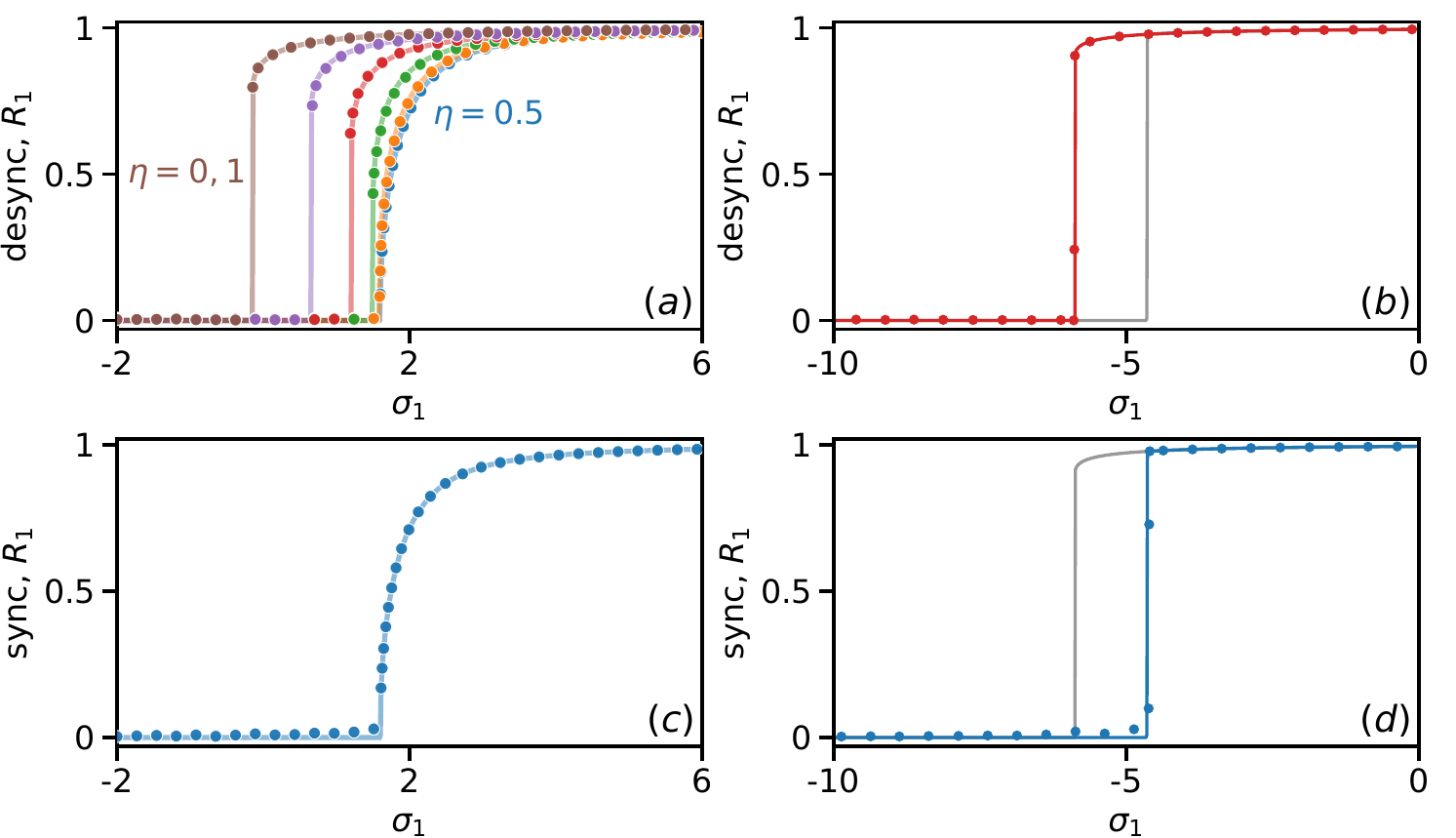}
\caption{\textbf{Numerical evidence of distinct critical behaviors for even and odd $h$ in Cup and Cap Synchronization with $M=2$.} The desynchronization (a,b) and synchronization (c,d) transitions are investigated by plotting the order parameter $R_1$ as a function of $\sigma_1$.
For even $h$, the desynchronization transition depends on the initial phase bias, while the synchronization transition is always continuous. In contrast, for odd $h$, the desynchronization transition is independent of the initial phase bias, while the synchronization transition can be discontinuous.
(a) Desynchronization transition for $h=4$ and $\sigma_2=3$, with
different initial phase biases $\eta$, defined by $P[\theta\in(-\pi,\pi)]=\eta$ and $P[\theta\in(\pi,3\pi)]=1-\eta$.
Results are shown for $\eta=1$ (brown), $0.9$ (purple),
$0.8$ (red), $0.7$ (green), $0.6$ (orange), and $0.5$ (blue).
(c) Synchronization transition for $h=4$, $\sigma_2=3$,
and $\eta=0.5$. (b,d) Desynchronization and synchronization transitions for $h=5$ and $\sigma_2=10$, respectively, revealing a stable hysteresis loop.
In all panels, numerical simulations (symbols) are in excellent
agreement with theoretical predictions (solid lines).
Results are obtained on a fully connected network of $N=10^4$
nodes with normally distributed intrinsic frequencies.}
    \label{fig:simulation}
\end{figure}
\subsection{Unified critical behavior of the Cup and Cap Synchronization including up to 3-body interactions}
The $M=2$ Cup and Cap Synchronization model provides a unified framework that displays critical behavior previously considered the outcome of distinct mechanisms reflected by different higher-order synchronization models. 
The critical behavior of the Cup and Cap Synchronization dynamics with $M=2$, Eq.~(\ref{dynamics_phase_M2}) is here investigated  when the intrinsic frequency distribution is a standard normal distribution, i.e. $\omega\sim \mathcal{N}(0,1)$ (see SI for the phase diagram corresponding to Lorentzian intrinsic frequency distribution). The case $h=1$ is unique, as  the dynamics reduces to the standard Kuramoto dynamics, upon redefinition of the coupling constant (as $f(x)=[\cos(\hat{\bm\eta}/2)]^{h-1}=1$).
Setting aside the case $h=1$, we observe significant differences among the critical behavior associated with   even values of $h$ and odd values of $h$. This different behavior  reflects the different periodicity of the dynamics: $4\pi$ for $h$ even and $2\pi$ for $h$ odd.
The phase diagram of the models obtained for $0\leq h\leq 5$ is shown in Fig.~\ref{fig:phasediagram}.

For $h$ even the synchronization transition reduces to that obtained for unbiased phase distribution, i.e. $\eta=0.5$ for which $X_{p/2}$ with $p$ odd is zero, i.e. $X_{p/2}=0$. This implies that for $h$ even  the synchronization reduces to the one of the standard (pairwise) Kuramoto model. This implies that the transition is always continuous and  that  the synchronization dynamics is exclusively driven by the coupling constant $\sigma_1\neq0$. Thus, for $\sigma_1=0$ there is no synchronization occurring as a function of $\sigma_2$. 

For $h$ even, the desynchronization transition can be discontinuous and abrupt if the initial condition is biased $\eta\neq 1/2$ with the desynchronization threshold depending on the bias $\eta$.

This scenario, observed for $h$ even, is reminiscent of the behavior observed for the higher-order synchronization proposed in Refs.~\cite{tanaka2011multistable,skardal2019abrupt}. The main difference is that the Cup and Cap Synchronization extends the periodicity of the dynamics from $2\pi$ of the standard Kuramoto model to $4\pi$, whereas in Refs.~\cite{tanaka2011multistable,skardal2019abrupt} the periodicity is reduced to $\pi$.
Another distinctive property of the Cup and Cap Synchronization dynamics  is that the higher-order dynamics is naturally coupled to the standard pairwise Kuramoto term without introducing further harmonics in Eq.~(\ref{eq:dyn_Lambda}). Consequently, the Cup and Cap Synchronization displays a synchronization transition whereas the original model in  Refs.~\cite{tanaka2011multistable,skardal2019abrupt} does not.

Moving from $h$ even to $h$ odd, the phase diagram changes significantly as a result of the restored $2\pi$ periodicity. With the exception of the case $h=1$ which, as mentioned before, reduces to the standard Kuramoto model albeit with a coupling constant depending on both $\sigma_1$ and $\sigma_2$, for $h$ odd the phase diagram is independent of the initial conditions. The $h$ odd (with $h\neq 1$) phase diagram  displays  continuous and discontinuous phase transitions with a thermodynamically stable hysteresis loop.
Therefore, for $h$ odd, the behavior of the Cup and Cap Synchronization model displays a critical behavior that is in the same universality class of the higher-order model proposed in Refs.~\cite{bick2016chaos,skardal2020higher}. Our theoretical predictions are perfectly matched with extensive numerical simulations for both $h$ even  and $h$ odd (see Fig.~\ref{fig:simulation}).
\begin{figure}[htbp]
    \includegraphics[width=0.9\columnwidth]{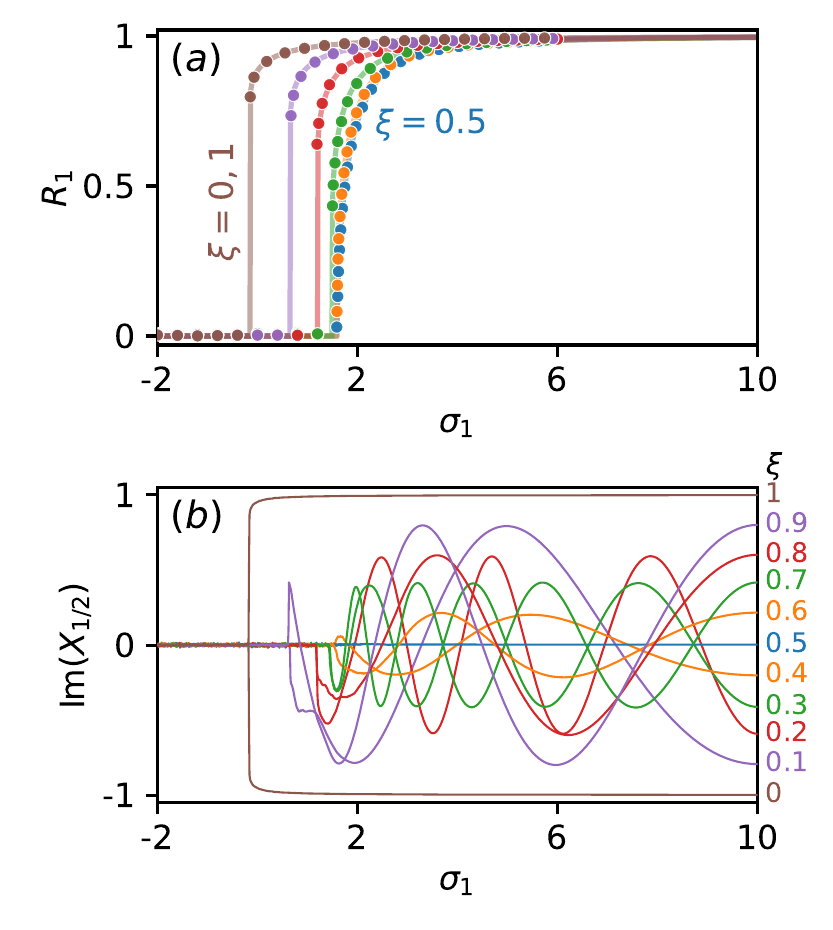}
    \caption{\textbf{Imaginary half-integer order parameters induced by a specific bias in the initial conditions of the phases  in Cup and Cap Synchronization with $M=2$ and even $h$.}
For even $h$, the initial phase bias $\xi$, defined by
$\mathbb{P}(\theta=\pi)=\xi$ and
$\mathbb{P}(\theta=3\pi)=1-\xi$, induces imaginary half-integer
order parameters during the desynchronization transition.
Panels (a) and (b) show the order parameter $R_1$ and the imaginary
part of the half-integer order parameter $\operatorname{Im}(X_{1/2})$,
respectively, as functions of the coupling strength $\sigma_1$,
for $h=4$ and $\sigma_2=3$.
For $\xi=0$ and $\xi=1$, the absence of phase frustration leads
to a stable imaginary half-integer order parameter, whereas
intermediate values of $\xi$ reveal the effects of phase frustration.
Different colours indicate different values of $\xi$, with
$\xi$ and $1-\xi$ represented by the same colour.
Simulations are performed on a fully connected network
of $N=10^4$ nodes with Gaussian intrinsic frequency
distribution.}
    \label{fig:sigma2posh4}
\end{figure}
These results show that the $M=2$ Cup and Cap Synchronization model is a unified approach to node-based synchronization showing how different critical behaviors previously thought to be irreconcilable can be recast within the same topological synchronization model.

\subsection{Imaginary half-integer order parameters}
For  $h$ even, the emergence of imaginary half-integer order parameters provides a further manifestation of the richer collective dynamics associated with the $4\pi$-periodic phase space, whose additional degrees of freedom are not resolved by conventional integer-order parameters.
Interestingly, we observe that in this scenario,  since the half-integer order parameters always occur in pairs in the dynamical equations,  it is also possible to bias the initial condition of the desynchronization dynamics to induce imaginary half-integer order parameters.
Specifically, in order to illustrate this phenomenon,  we consider the case $M=2$ with $h=h_1$ even, and an initial condition 
$\theta_{[i]}=\pi$ with probability $\xi$ and $\theta_{[i]}=3\pi$ with probability $1-\xi$. These initial conditions force $\cos(\theta)<0$ and imaginary half-integer order parameters, such as $X_{1/2}=\textrm{i}\sum_{j=1}^N\sin(\theta_{[j]}/2)$.  Interestingly, the order parameter $X_1$ during the desynchronization transition can be predicted to follow the same behavior as the desynchronization transition observed for real half-integer order parameters with  bias $\eta$ with $\xi=\eta$ (see Methods for details).
The numerical results for $M=2$ and $h=4$ obtained for different values of the bias $\xi$ are shown in Fig.~$\ref{fig:sigma2posh4}$ to behave as predicted by our theory.
For $\xi=0$ and $\xi=1$ we observe that during the desynchronization, the half-integer parameters remain imaginary and unfrustrated, while for intermediate values of $\xi$, the imaginary part of the half-integer order parameters fluctuates and is more unstable.

\begin{table}
\begin{center}
\begin{tabular}{ ||c|c|c|c|| } 
 \hline
 \hline
 Parity & $\Theta^{(1)}$ & $\Theta^{(2)}$ & $\Theta^{(3)}$ \\
 \hline
 $h_1$ and $h_2$ both are odd:  &$X_1$ &$X_\mathbb{Z}X_\mathbb{Z}$ & $X_\mathbb{Z}X_\mathbb{Z}X_\mathbb{Z}$ \\ 
 $h_1$ and $h_2$ both are even:  &$X_1$ &$X_{\mathbb{Z}/2}X_{\mathbb{Z}/2}$ & $X_{\mathbb{Z}/2}X_\mathbb{Z}X_{\mathbb{Z}/2}$ \\
 $h_1$ even and $h_2$ odd:  &$X_1$ &$X_{\mathbb{Z}/2}X_{\mathbb{Z}/2}$ & $X_{\mathbb{Z}/2}X_{\mathbb{Z}/2}X_\mathbb{Z}$ \\ 
 $h_1$ odd and $h_2$ even:  &$X_1$ &$X_\mathbb{Z}X_\mathbb{Z}$ & $X_{\mathbb{Z}}X_{\mathbb{Z}/2}X_{\mathbb{Z}/2}$ \\ 
 \hline
 \hline
\end{tabular}
\end{center}
\caption{\textbf{Characterization of the interaction terms in Cup and Cap Synchronization with $M=3$.} The contributions of two-body, three-body, and four-body interactions are investigated as functions of the parity of $h_1$ and $h_2$ by emphasizing the structure of  $\Theta^{(1)}$, $\Theta^{(2)}$, and $\Theta^{(3)}$,
which enter the self-consistency equation for $\Lambda$,
in terms of integer and half-integer order parameters.
The case of odd $h_1$ and even $h_2$ emerges as particularly
relevant, as the three interaction terms give rise to distinct
contributions to the synchronization dynamics, suggesting
a dependence on the initial phase bias coexisting with 
a stable hysteresis loop leading to a critical behavior that has no equivalence in models that include only interactions mediated by nodes, edges and triangles ($M=2$).}
\label{table1}
\end{table}

\subsection{Inclusion of 4-body interactions}
Having shown that  for $M=2$, the Cup and Cap Synchronization  is able to unify critical behavior of higher-order node-based dynamics previously thought to arise from different mechanisms, the pressing open question concerns the consequences of including interactions of even higher order. By extending the model to $M=3$, here we show that the Cup and Cap Synchronization model displays an unexpected critical behavior that has no equivalent in higher-order node-based synchronization models mediated only by nodes, edges and triangles.
The $M=3$ Cup and Cap Synchronization model reads, 
\bea
  \frac{d\theta_{[i]}}{dt} &=& \omega_{[i]}+\frac{\sigma_1}{N}\sum_{j=1}^N a_{[ij]}W_{[ij]}+\frac{\sigma_2}{N^2}\sum_{j=1}^N\sum_{k=1}^Na_{[ijk]}W_{[ijk]}\nonumber \\
   &&+\frac{\sigma_3}{N^3}\sum_{j=1}^N\sum_{k=1}^N\sum_{l=1}^Na_{[ijkl]} W_{[ijkl]}
   \label{eq:dynamics_phase_M3}
   \eea 
   where $W_{[ij]}$, $W_{[ijk]}$  and $W_{[ijkl]}$ are given by
   \bea
   W_{[ij]}&=& \sin{(\theta_{[j]}-\theta_{[i]})}\nonumber\\
   W_{[ijk]}&=&\sin(\theta_{[j]}-\theta_{[i]})f_1(\eta_{[jk]}/2). \\
   W_{[ijkl]}&=&\sin(\theta_{[j]}-\theta_{[i]})f_1(\eta_{[jk]}/2)f_2(\eta_{[kl]}/2).\nonumber
   \eea
Choosing $f_{r}(x)=[\cos(x)]^{h_r-1}$, by  analysing the corresponding trigonometric relations, as for $M=2$, (see Supporting Information (SI) for details), the $M=3$ Cup and Cap Synchronization dynamics reduces to 
\bea
\dot{\theta}_{[i]} = \omega_{[i]} - \text{Im}\left[{\Lambda} \exp(\textrm{i}\theta_{[i]})\right]
\label{eq:dyn_Lambda4}
\eea
with 
\bea
\Lambda &=&\left(\sigma_1 \Theta^{(1)} + \frac{\sigma_2}{2^{h_1}}\Theta^{(2)}  + \frac{\sigma_3}{2^{h_1+h_2-2}}\Theta^{(3)}\right)^*\eea
where $\Theta^{(1)}=X_1$, $\Theta^{(2)}$ is defined in Eq.~(\ref{eq:Theta2}) and where $\Theta^{(3)}$ is given by 
\bea
\Theta^{(3)}&=&\sum_{n=0}^{h_1-1}\sum_{m=0}^{h_2-1}\binom{h_1-1}{n}\binom{h_2-1}{m}  \left[X_{(1-2n+h_1)/2} \right.\nonumber \\
&& \left. X_{(2n-2m-h_1+h_2)/2} X_{(2m-h_2+1)/2}\right].
\eea
\begin{figure}
    \includegraphics[width=0.9\columnwidth]{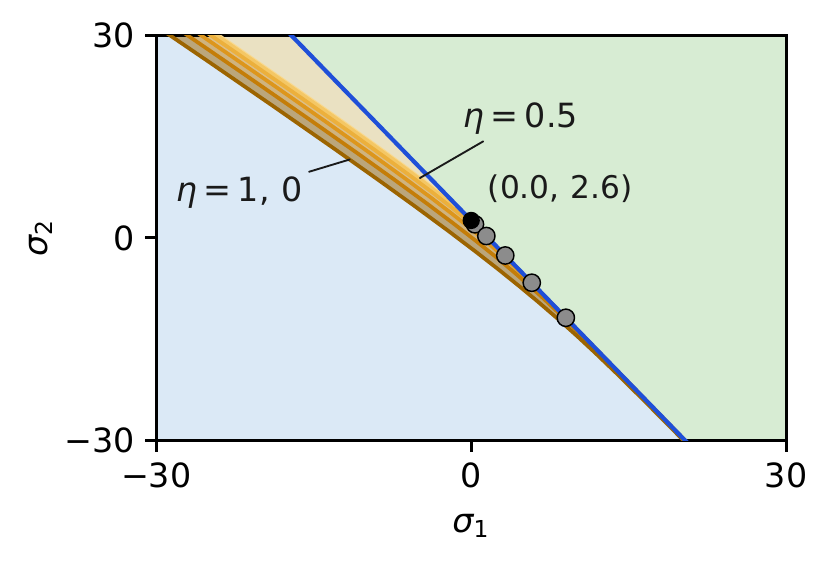}
    \caption{\textbf{Four-body interactions in Cup and Cap Synchronization give rise to critical behavior absent in higher-order synchronization models restricted to nodes, edges, and triangles.}
The phase diagram of Cup and Cap Synchronization with $M=3$,
odd $h_1$, and even $h_2$ reveals the coexistence of a stable
hysteresis loop and a dependence of the desynchronization transition on the initial phase bias $\eta$. The phase diagram is shown in the $(\sigma_1,\sigma_2)$ plane for $h_1=5$, $h_2=2$, and $\sigma_3=5$, with Gaussian-distributed intrinsic frequencies $g(\omega)$. The blue, green, and brown regions indicate the incoherent, synchronized, and bistable states, respectively. The blue line marks the subcritical pitchfork bifurcation, while the brown lines, from light to dark, indicate the saddle-node bifurcations for $\eta$ or $1-\eta=0.5,0.6,0.7,0.8,0.9,1$,  respectively. The black dot marks the cusp point at $(\sigma_1^C,\sigma_2^C)=(0,2.6)$, 
while the gray dots indicate the onset of continuous  desynchronization transitions for $|\eta-0.5|<\delta^{\star}(\sigma_2)$. These occur at $(\sigma_1,\sigma_2)$ given by $(0.36,1.98)$ (for $\delta^{\star}= 0.0997$), $(1.44,0.25)$ (for $\delta^{\star}=0.1998$), $(3.24, -2.64)$ (for $\delta^{\star}=0.3000$), $(5.77,-6.68)$ (for $\delta^{\star}=0.4001$), and $(9.01,-11.87)$ (for $\delta^{\star}=0.5000$). }
    \label{fig:phasediagM3}
\end{figure}

\begin{figure}[!ht!]
    \includegraphics[width=0.8\columnwidth]{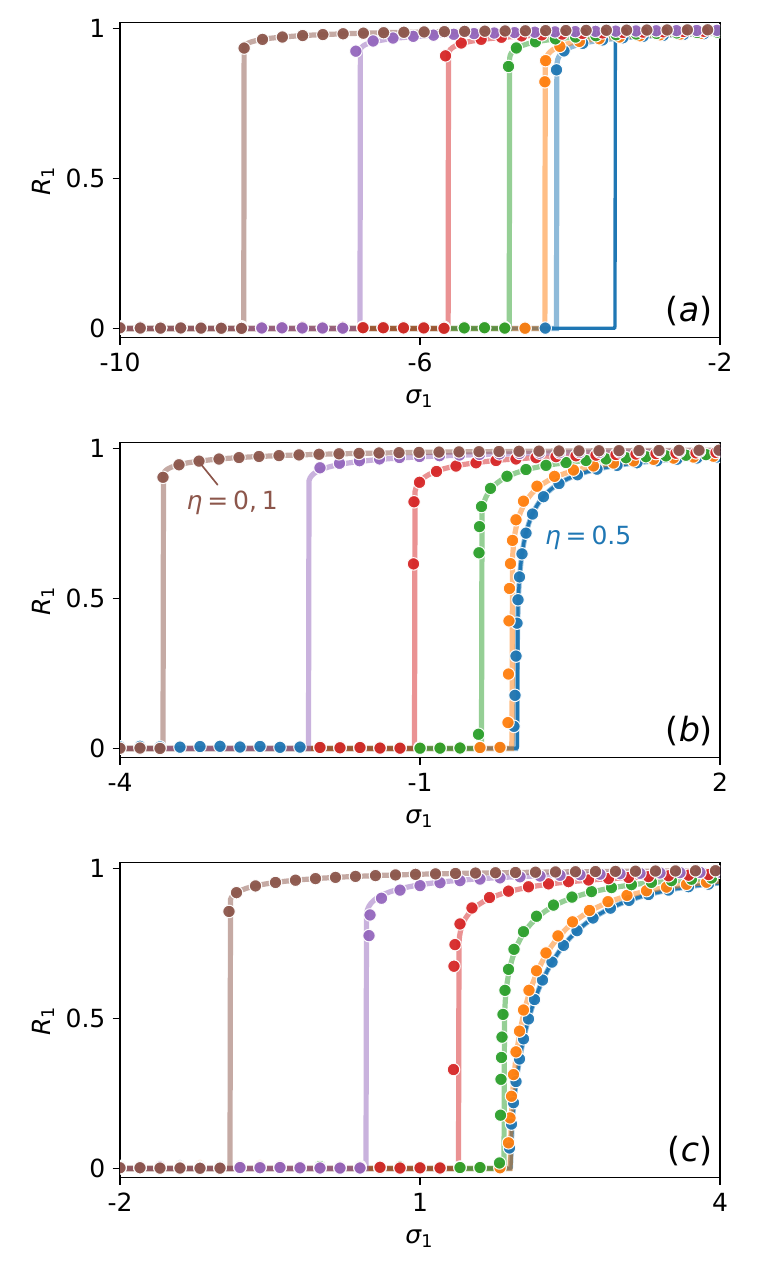}
    \caption{\textbf{Numerical evidence that four-body interactions in Cup and Cap Synchronization can fundamentally change the critical behavior of the model.} For $M=3$, odd $h_1$ and even $h_2$, the synchronization dynamics displays the coexistence of hysteresis and a strong dependence of the desynchronization transition on the initial phase bias $\eta$. The order parameter $R_1$ is plotted as a function of $\sigma_1$ for $h_1=5$, $h_2=2$, $\sigma_3=5$, and three values of $\sigma_2$ above, at, and below the cusp $(\sigma_1^C,\sigma_2^C)=(0,2.6)$ shown in Fig.~\ref{fig:phasediagM3}. (a) For $\sigma_2=8>\sigma_2^C$, the synchronization transition (blue line) and the unbiased desynchronization transition ($\eta=0.5$, blue line) form a stable hysteresis loop, while biased desynchronization transitions remain abrupt. (b) At the cusp, $\sigma_2=2.6=\sigma_2^C$, the synchronization and desynchronization transitions coincide and become continuous for $\eta=0.5$, while biased desynchronization transitions remain discontinuous. (c) Below the cusp, $\sigma_2=-0.5<\sigma_2^C$, a critical bias $\delta^\star=\delta^\star(\sigma_2)$ separates continuous and discontinuous desynchronization transitions. For $|\eta-0.5|<\delta^\star$, the transition is continuous and occurs at the same threshold as for $\eta=0.5$, although the order parameter $R_1$ takes larger values above the transition. For $|\eta-0.5|>\delta^\star$, the desynchronization transitions remain abrupt. Colors indicate $\eta$ (or $1-\eta$): $1$ (brown), $0.9$ (purple), $0.8$ (red), $0.7$ (green), $0.6$ (orange), and $0.5$ (blue).  Numerical simulations are in excellent agreement with theoretical predictions (solid lines) obtained  on a fully connected network of $N=10^4$ nodes with Gaussian intrinsic frequency distributions.
}
    \label{fig:simulationM3}
\end{figure}
The expression of $\Lambda$ allows us to explore the role of including four-body interactions in the Cup and Cap Synchronization model. We are specifically interested in classifying $\Theta^{(1)},\Theta^{(2)},\Theta^{(3)}$ according to  their composition in terms of the integer and half-integer order parameters as a function of the parity of $h_1$ and $h_2$.
Table~\ref{table1} reveals that introducing four-body interactions leads to an unprecedented scenario for the choice of $h_1$ odd and $h_1\neq 1$ and $h_2$ even. Indeed, in this case $\Theta^{(1)}$ is responsible for the pairwise, standard Kuramoto-like term, $\Theta^{(2)}$ is responsible for a three-body interaction term leading to a stable hysteresis loop, while  $\Theta^{(3)}$, expressed in terms of a pair of half-integer order parameters and one integer order parameter,  demonstrates that  the further introduction of four-body interactions can lead to an additional  dependence on the initial condition of the desynchronization transition.
The phase diagram corresponding to $M=3$ for $\sigma_3=5$, with $h_1$ odd and $h_2$ even (specifically $h_1=5,h_2=2$) is shown in  Fig.~$\ref{fig:phasediagM3}$ revealing a critical behavior that combines a stable hysteresis loop with a strong dependence on the bias $\eta$ on the initial condition in the desynchronization transition.
To further illustrate this critical behavior in Fig.~\ref{fig:simulationM3} we present the synchronization and desynchronization transition as functions  of $\eta$, observed for the same model, and for different values of $\sigma_2$ above, at and below the cusp $\sigma_2^C$, i.e. with $\sigma_2>\sigma_2^{C},\sigma_2=\sigma_2^{C},$ and $\sigma_2<\sigma_2^{C}$.
For $\sigma_2>\sigma_2^{C}$ the stable hysteresis loop is observed for $\eta=0.5$  with the desynchronization transitions corresponding to $\eta\neq 1/2$ remaining abrupt.
For $\sigma_2=\sigma_2^{C}$, the unbiased ($\eta=0.5$) synchronization transition displays a cusp. Thus, the synchronization and desynchronization coincide for the unbiased dynamics.
For $\sigma_2<\sigma_2^{C}$ the $\eta=0.5$ transition is continuous while the biased desynchronization transitions have different behaviors depending on the value of $\eta$. Indeed, the model displays a set of thresholds $\delta^{\star}=\delta^{\star}(\sigma_2)$ that determine distinct critical behaviors for the desynchronization transitions. For $|\eta-0.5|<\delta^{\star}(\sigma_2)$ the desynchronization transitions are continuous with the same synchronization threshold as for $\eta=0.5$ but displaying a larger order parameter $R_1$ above the threshold. However, for $|\eta-0.5|>\delta^{\star}(\sigma_2)$ the desynchronization transitions remain abrupt and discontinuous.
The points  $(\sigma_1^{\star},\sigma_2^{\star})$ where a given desynchronization transition with bias $\eta$ becomes continuous are indicated as gray dots in the phase diagram shown in Figure $\ref{fig:phasediagM3}$.
Summing up, the introduction of  four-body interactions in the Cup and Cap Synchronization model leads to a critical behavior that has no equivalent in models where higher-order interactions are only mediated by nodes, edges, and triangles, demonstrating the richness of this unified model of node-based higher-order synchronization.

\section{Discussion}

Despite significant advances in higher-order synchronization, understanding how interactions mediated by higher-order simplices shape the critical behavior of collective dynamics remains a fundamental challenge. In particular, a central research question is  whether the distinct critical behaviors observed in higher-order node-based synchronization originate from fundamentally different mechanisms or can be understood within a unified theoretical framework.

In this work, we address this challenge by introducing Cup and Cap Synchronization as a unifying algebraic topological framework for higher-order node-based dynamics. By expressing the dynamics in terms of cup and cap products, which lift and lower topological signals across arbitrary dimensions, the framework provides a theoretical foundation for describing node-based synchronization mediated by higher-order interactions.

Our results demonstrate that critical behaviors previously thought to originate from distinct synchronization mechanisms can emerge within the same theoretical framework, revealing a common topological foundation underlying different classes of higher-order synchronization. Beyond this unifying perspective, Cup and Cap Synchronization reveals novel dynamical phenomena that have no equivalent in previously proposed models. Already at the three-body level, the model reveals a richer collective dynamics associated with its $4\pi$-periodicity, characterized by half-integer order parameters that can become imaginary depending on the initial phase bias. These additional order parameters capture collective information that is inaccessible to conventional integer-order parameters. Remarkably, the inclusion of four-body interactions gives rise to a distinct critical behavior characterized by the coexistence of a stable hysteresis loop and a strong dependence on initial conditions along the desynchronization branch. This demonstrates that interactions mediated by higher-dimensional simplices can fundamentally alter the critical properties of synchronization, rather than merely introducing quantitative corrections to existing models.

More broadly, our findings establish algebraic topology as a general theoretical foundation for understanding higher-order node-based collective dynamics, extending its role beyond the previously investigated synchronization of higher-order topological signals. Looking forward, the Cup and Cap Synchronization is a unified theoretical framework that opens new directions for investigating how the interplay between topology and interactions of increasing order shapes collective behavior, potentially revealing further interplay between topology and nonlinear dynamics.
\section*{Methods}
\subsection*{Limiting cases of the $M=2$ Cup and Cap Synchronization}
In this Methods section we reveal the relation of the $M=2$ Cup and Cap Synchronization with existing models \cite{tanaka2011multistable,skardal2019abrupt,bick2016chaos,skardal2020higher} of higher-order node-based synchronization dynamics.
For $h_1=h=0$, the Cup and Cap Synchronization dynamics includes a three-body interaction term of the type 
\bea
W_{[ijk]}=\sin((\theta_{[k]}+\theta_{[j]}-2\theta_{[i]})/2)\eea and a pairwise term $W_{[ij]}=\sin(\theta_{[j]}-\theta_{[i]})$. Thus, in the absence of the pairwise term, i.e. for $\sigma_{1}=0$, and upon identification of $\bm\phi=\bm\theta/2$, the Cup and Cap Synchronization dynamics can be recast into the higher-order node-based model with three body term  $\sin(\phi_{[k]}+\phi_{[j]}-2\phi_{[i]})$\cite{tanaka2011multistable,skardal2019abrupt}. Note however that the Cup and Cap Synchronization model with $\sigma_1\neq0 $ cannot be recast in previously proposed higher-order synchronization models.
For $h_1=h=3$, the Cup and Cap Synchronization dynamics can be related to the higher-order node-based synchronization with three body term  $\sin(2\phi_{[k]}-\phi_{[j]}-\phi_{[i]})$ as we have 
\bea
W_{[ijk]}&=&\frac{1}{8}\left[3\sin(\theta_{[j]}-\theta_{[i]})+3\sin(\theta_{[k]}-\theta_{[i]})\right.\nonumber \\
 &&\hspace{-5mm}\left.+\sin(2\theta_{[j]}-\theta_{[k]}-\theta_{[i]})+\sin(2\theta_{[k]}-\theta_{[j]}-\theta_{[i]})\right].
\eea
Thus, modulus rescaling of the coupling constants, this model can be recast into the model proposed in Refs.~\cite{bick2016chaos,skardal2020higher}, for every value of  $\sigma_1$ and $\sigma_2$. 
\subsection*{Imaginary half-integer order parameters (case $M=2$)}
In this section  we provide further evidence of the existence of  dynamical states for Cup and Cap synchronization in which the half-integer order parameters are imaginary.
These states occur in the desynchronization transitions, where the initial phases are biased via the parameter $\xi$ that determines the probability that the initial phases are equal to $\pi$ whereas $1-\xi$ indicates the probability that the initial phases are equal to $3\pi$. Here we show that the dynamics under these conditions is equivalent to the one when the phases are initially biased with probability $\eta$ to be $0$ and with probability $1-\eta$ to be $2\pi$.
The initial configuration where the bias is given by $\xi$ is obtained from the one in which $\eta=\xi$ by performing the transformation \bea
\theta\to \theta+\pi.
\eea
Under this transformation, we have 
\bea
\cos(\theta)\to -\cos(\theta),\quad \sin(\theta)\to -\sin (\theta),\nonumber \\
\cos(\theta/2)\to -\sin(\theta/2),\quad \sin(\theta/2)\to \cos(\theta/2).
\eea
Therefore, the integer order parameters $X_k$ and the half-integer order parameters $X_{p/2}$ with $p$ odd, transform as 
\bea
X_k=R_k&\to& (-1)^kR_k,\nonumber \\
X_{p/2}&\to& \textrm{i}X_{p/2}.
\eea
Therefore owing to the gauge symmetry, $\Lambda$ can always be chosen real, and the Cup and Cap Synchronization equation becomes
\bea
\Lambda=-\sigma_1\Theta^{(1)}-\frac{\sigma_{2}}{2^h}\Theta^{(2)}.
\eea
Since $\Theta^{(1)},\Theta^{(2)}$ do not depend on the sign of $\Lambda$, by considering the transformation 
\bea
\Lambda\to -\Lambda\eea
we recover the self-consistent equation for 
\bea
\Lambda=\sigma_1\Theta^{(1)}+\frac{\sigma_{2}}{2^h}\Theta^{(2)}.
\eea
It follows that  the model which is biased to have phases around $\pi$ and $3\pi$ with bias given by $\xi$
has the same phase diagram of the standard model with the choice $\xi=\eta$ as shown in Figure \ref{fig:sigma2posh4}.
\subsection*{Details on the numerical integration of the Cup and Cap Synchronization model}
In this work, the Cup and Cap Synchronization dynamics  is integrated numerically using the fourth-order Runge-Kutta method with time step $\Delta t=0.005$. The synchronization and desynchronization transitions are numerically investigated by raising and lowering adiabatically the values of the coupling constant $\sigma_1$. The  bias in the initial conditions of the desynchronization transition is implemented in Figures \ref{fig:simulation} and \ref{fig:simulationM3} by selecting at random a fraction $\eta$ of the nodes whose phases are drawn from a uniform distribution  $U((-\pi,\pi])$ while the remaining $1-\eta$ fraction of the nodes have phases drawn from a uniform distribution  $U((\pi,3\pi])$. The bias that induces imaginary half-integer order parameters as shown in Figure \ref{fig:sigma2posh4} is implemented by selecting at random a fraction $\xi$ of the nodes whose phases are $\theta=\pi$ while the remaining $1-\xi$ fraction of the nodes have phases $\theta=3\pi$. All simulations are performed on fully connected networks with $N=10^4$ and adiabatic steps $\delta \sigma=0.005$. The equilibration time is chosen to be $T_{\textrm{eq}}=15$ and the values of the order parameters are averaged over times $T_{\textrm{eq}} \leq t \leq T_{\textrm{max}}$ with $T_{\textrm{max}}=20$.
\bibliography{bibliography}


\renewcommand\theequation{{S-\arabic{equation}}}
\renewcommand\thetable{{S-\Roman{table}}}
\renewcommand\thefigure{{S-\arabic{figure}}}
\setcounter{equation}{0}
\setcounter{figure}{0}
\setcounter{section}{0}

\onecolumngrid
\vspace{0.7cm}
\begin{center}
{\Large\bf             SUPPLEMENTARY INFORMATION}
\end{center}

\section{Derivation of the trigonometric relation}

In this section we provide the derivation of the trigonometric relations that turn the $2$-dimensional and the $3$-dimensional coupling terms of the Cup and Cap synchronization on fully connected networks.
The Cup and Cap Synchronization dynamics reads, for $M=3$, as
\bea
\frac{d{\bm\theta}}{dt}={\bm\omega}+\sum_{n=1}^3 \frac{\sigma_n}{N^n} \left[(\hat{\bm\chi}^{(n-1)}\smile\sin(\hat{\bm\eta}))\frown{\bm \alpha}^{(n)}\right],
\label{eq:ccsS}
\eea
where we assume for $n>0$
\bea
\hat{\bm\chi}^{(0)}&=&{\bf 1},\nonumber \\
\hat{\bm\chi}^{(1)}&=&f_{1}(\hat{\bm\eta}/2),\nonumber \\
\hat{\bm\chi}^{(2)}&=&f_{2}(\hat{\bm\eta}/2)\smile f_1(\hat{\bm\eta}/2),
\eea
where in this work we always assume
 \bea
 f_1(x)=(\cos x)^{h_1-1},\quad f_2(x)=(\cos x)^{h_2-1}.
 \eea
 Element-wise the Cup and Cap synchronization dynamics reads:
\bea
   \frac{d\theta_{[i]}}{dt} &=& \omega_{[i]}+\frac{\sigma_1}{N}\sum_{j=1}^N a_{[ij]} \sin{(\theta_{[j]}-\theta_{[i]})}\nonumber \\&&+\frac{\sigma_2}{N^2}\sum_{j=1}^N\sum_{k=1}^Na_{[ijk]}\sin(\theta_{[j]}-\theta_{[i]})f_1(\eta_{[jk]}/2)\nonumber \\
   &&+\frac{\sigma_3}{N^3}\sum_{j=1}^N\sum_{k=1}^N\sum_{l=1}^Na_{[ijkl]}\sin(\theta_{[j]}-\theta_{[i]})f_1(\eta_{[jk]}/2)f_2(\eta_{[kl]}/2) 
\eea
These dynamical equations can be recast in the following equations depending on the global order parameter $\Lambda$, as follows:
\bea
\dot{\theta_{[j]}}=\omega_{[j]}-\mbox{Im}\left[\Lambda e^{\textrm{i}\theta_{[j]}}\right].
\eea
Here $\Lambda$ is expressed in terms of the coupling constants $\sigma_n$ and the complex order parameters $X_{p/2}$ of the theory  given by 
\bea
 {X}_{p/2} ={R}_{p/2}e^{\textrm{i}\Psi_{p/2}}=\frac{1}{N} \sum_{j=1}^N e^{\textrm{i}p \theta_{[j]}/2}.\label{eq:X}
\eea

In the following we give the details for the derivation of the $\Lambda$ term when only $3$-body interactions are included and when also $4$-body interactions are included.
\subsection{The  $2$-dimensional coupling term mediated by triangles}
The $2$-dimensional coupling term of the fully connected Cup and Cap synchronization is expressed in terms of the 1-cochain $\hat{\bm\eta}\in C^1$ which is the coboundary of the node phases  $\hat{\bm\theta}$, i.e. $\hat{\eta}=\delta\hat{\bm\theta}$ and an element-wise function $f_1((\hat{\bf{\eta}}/2))$ with $f_1(x)=(\cos x)^{h_1-1}$. 
Specifically, the term reads
\bea
\left(f_{1}(\hat{\bm{\eta}}/2)\smile \sin (\hat{\bm\eta})\right)\frown {\bm\alpha}^{(2)}=\sum_{j=1}^N\sum_{k=1}^Na_{[ijk]}\left( \cos{\frac{\theta_{[k]}-\theta_{[j]}}{2}}\right)^{h_1} \sin{\left( \frac{\theta_{[k]} + \theta_{[j]} - 2\theta_{[i]}}{2}\right)}{\bf e}_{[i]}
\eea
The binomial expansion of the cosine function gives
\bea  
    \left(\cos\left(\frac{\theta_{[k]}-\theta_{[j]}}{2}\right)\right)^{h_1} &=& \frac{1}{2^h}\sum_{n=0}^{h_1}\binom{h_1}{n} \exp\left(\textrm{i}(2n-h_1)\frac{\theta_{[k]}-\theta_{[j]}}{2}\right),\nonumber \\
    \sin\left(\frac{\theta_{[k]}+\theta_{[j]}- 2\theta_{[i]}}{2}\right) &=& \frac{1}{2\textrm{i}}\left(\exp\left(\i\frac{\theta_{[k]}+\theta_{[j]}- 2\theta_{[i]}}{2}\right)- \exp\left(-\i\frac{\theta_{[k]} + \theta_{[j]} - 2\theta_{[i]}}{2}\right)\right).
    \label{binomial_expansion}
\eea
By multiplying the two terms and rearranging, we get 
\bea
&&\frac{1}{N^2}\sum_{j=1}^N\sum_{k=1}^N\left(\cos\left(\frac{\theta_{[k]}-\theta_{[j]}}{2}\right)\right)^{h_1}\sin\left( \frac{\theta_{[k]} + \theta_{[j]} - 2\theta_{[i]}}{2}\right) \nonumber \\
&&= \frac{1}{2^{h}}\frac{1}{N^2}\sum_{j=1}^N\sum_{k=1}^N\sum_{n=0}^{h_1} \binom{h_1}{n} \sin \left(\frac{2n-h_1+1}{2} \theta_{[k]} - \frac{2n -h_1-1}{2} \theta_{[j]} - \theta_{[i]} \right)\nonumber\\&&=\frac{1}{2^{h_1}}\sum_{n=0}^{h_1} \binom{h_1}{n} \mbox{Im}\left[X_{(2n-h_1+1)/2}X_{(-2n+h_1+1)/2}e^{-\textrm{i}\theta_{[i]}}\right]\nonumber \\
\eea

\subsection{The $3$-dimensional coupling term mediated by tetrahedra }

We now wish to express the $4$-body interaction term in terms of the order parameters. This term reads, 
\bea
\left[(f_{2}(\hat{\bm\eta}/2)\smile f_1(\hat{\bm\eta}/2))\smile\sin(\hat{\bm\eta})\right]\frown{\bm \alpha}^{(3)}=
\sum_{i=1}^N\sum_{j=1}^N\sum_{k=1}^N\sum_{l=1}^Na_{[ijkl]}\sin(\theta_{[j]}-\theta_{[i]})f_1(\eta_{[jk]}/2)f_2(\eta_{[kl]}/2){\bf e}_{[i]}
\eea
Focusing on the expression for the element associated with node $i$, we consider the trigonometric expansions, leading to the desired expression as a function of the order parameters $X_{p/2}$, i.e.
\bea
   &&\sin(\theta_{[j]}-\theta_{[i]})f_1(\eta_{[jk]}/2)f_2(\eta_{[kl]}/2)\nonumber\\
   &=& \frac{1}{2^{h_1 + h_2-1}\textrm{i}}\sum_{n=0}^{h_1-1}\sum_{m=0}^{h_2-1}\binom{h_1-1}{n}\binom{h_2-1}{m} \left(e^{\textrm{i}(
   \theta_{[j]}-\theta_{[i]}
   )} - e^{-\textrm{i}(\theta_{[j]}-\theta_{[i]})}\right) e^{\textrm{i}(2n-h_1+1)\frac{\theta_{[k]}-\theta_{[j]}}{2} + \textrm{i}(2m-h_2+1)\frac{\theta_{[l]}-\theta_{[k]}}{2}}\nonumber\\
   &=& \frac{1}{2^{h_1 + h_2-1}\textrm{i}}\sum_{n=0}^{h_1-1}\sum_{m=0}^{h_2-1}\binom{h_1-1}{n}\binom{h_2-1}{m} \sin\left(-\theta_{[i]} + \frac{1-2n+h_1}{2}
   \theta_{[j]} + \frac{2n-2m-h_1+h_2}{2}\theta_{[k]} + \frac{2m-h_2+1}{2}\theta_{[l]}\right).\nonumber 
   \eea
   Thus, we have that 
   \bea
   &&\frac{1}{N^3}\sum_{j=1}^N\sum_{k=1}^N\sum_{l=1}^N\sin(\theta_{[j]}-\theta_{[i]})f_1(\eta_{[jk]}/2)f_2(\eta_{[kl]}/2)\nonumber \\
   &=& \frac{1}{2^{h_1 + h_2-2}}\sum_{n=0}^{h_1-1}\sum_{m=0}^{h_2-1}\binom{h_1-1}{n}\binom{h_2-1}{m} \mbox{Im}\left[ X_{(1-2n+h_1)/2} X_{(2n-2m-h_1+h_2)/2} X_{(2m-h_2+1)/2} e^{-\textrm{i}\theta_{[i]}}\right]
\eea

\subsection{Expression for $\Lambda$}

From the above trigonometric expansions it follows that in the infinite network limit of fully connected networks for given $h_1,h_2$, the Cup and Cap Synchronization Dynamics can be written as 
\bea
\dot{\theta_{[i]}} = \omega_{[i]} - \text{Im}\left[\Lambda e^{\textrm{i}\theta_{[i]}}\right]
\eea
where
\bea
\Lambda^* &=& \sigma_1 X_1 + \frac{\sigma_2}{2^{h_1}}\sum_{n=0}^{h_1} \binom{h_1}{n} 
X_{(2n-h_1+1)/2}X_{(-2n+h_1+1)/2} \nonumber\\
&& + \frac{\sigma_3}{2^{h_1+h_2-2}}\sum_{n=0}^{h_1-1}\sum_{m=0}^{h_2-1}\binom{h_1-1}{n}\binom{h_2-1}{m}  X_{(1-2n+h_1)/2} X_{(2n-2m-h_1+h_2)/2} X_{(2m-h_2+1)/2}.
\eea
Since, as we will show below, the order parameters $X_{p/2}$ are functions of $\Lambda$, and $\Lambda$ can be chosen to be real,  this self-consistent set of equations can be written as 
\bea
\Lambda=f(\Lambda;\{\sigma_n\}).
\eea
where 
\bea
f(\Lambda;\{\sigma_n\})&=&\sigma_1 X_1 + \frac{\sigma_2}{2^{h_1}}\sum_{n=0}^{h_1} \binom{h_1}{n} X_{(2n-h_1+1)/2}X_{(-2n+h_1+1)/2} \nonumber\\
&& + \frac{\sigma_3}{2^{h_1+h_2-2}}\sum_{n=0}^{h_1-1}\sum_{m=0}^{h_2-1}\binom{h_1-1}{n}\binom{h_2-1}{m}  X_{(1-2n+h_1)/2} X_{(2n-2m-h_1+h_2)/2} X_{(2m-h_2+1)/2}.
\eea
\section{Order parameters}

\subsection{Integer order parameters $R_k$ with $k\in \{1,2,3\}$}
The integer complex order parameters $X_k=X_{-k}^*$ of the dynamical system defined by Eq.(\ref{eq:X})
are derived by using the Ott-Antonsen ansatz on the even part of the probability density for the phases.
For $k\in \{1,2,3\}$ we have that $R_k=|X_k|$ is given by
\bea
R_{1}&=&\int d\omega g(\omega) [c^{\star}(\omega)]^2=\int_{|\omega|\leq\Lambda} d\omega g(\omega) \sqrt{1 - \left(\frac{\omega}{\Lambda}\right)^2}\nonumber\\
R_2&=&\int_{|\omega|\leq\Lambda} d\omega g(\omega)\left[1-2\left(\frac{\omega}{\Lambda}\right)^2\right] +\int_{|\omega|>\Lambda} d\omega g(\omega) \left[ 1-2\left(\frac{\omega}{\Lambda}\right)^2 + 2\left |\frac{\omega}{\Lambda}\right |\sqrt{\left(\frac{\omega}{\Lambda}\right)^2 - 1} \right]\nonumber \\
R_3&=&\int_{|\omega|\leq\Lambda} d\omega g(\omega) \left[1-4\left(\frac{\omega}{\Lambda}\right)^2\right]\sqrt{1 - \left(\frac{\omega}{\Lambda}\right)^2}
\eea
While for Gaussian intrinsic frequency distribution $g(\omega)$ one needs to perform numerically these integrals for the Lorentzian intrinsic frequency distribution the integer order parameters admit a closed form expression derived starting from the expression of these order parameters in terms of Cauchy integrals.
Therefore, for a Lorentzian distribution of the intrinsic frequencies
\bea
g(\omega)=\frac{1}{\pi} \frac{\Delta}{(\omega^2 + \Delta^2)},
\label{Lorentzian}
\eea
the integer order parameters $R_k$ with $k\in \mathbb{Z}$ read
\bea
R_{k}= R_1^{|k|}\quad \mbox{with}\quad
R_1=\frac{\sqrt{\hat{\Lambda}^2 + \Delta^2}-\Delta}{\hat{\Lambda}}.
\eea
Indeed using the Ott-Antonsen notation explained in the main text we can express these integer order parameters in terms of $c^\star (\omega)$, and we obtain using the Cauchy integral, for $k>0$
\bea
 R_{k}=R_{-k}=\int d\omega g(\omega)[c^\star (\omega)]^{2k} = 2\pi \mathrm{i}  \frac{1}{2\pi \mathrm{i}}\left(c^\star(\mathrm{i}\Delta)\right)^{2k} = \left(c^\star(\mathrm{i}\Delta)\right)^{2k} = R_1^{|k|}.
 \eea

\subsection{Half-integer order parameters $X_{p/2}$ with $p\in \{1,3,5\}$}

As explained in the main text, the Ott-Antonsen ansatz does not apply to the odd part of the probability density of the phases, so the half-integer order parameters require the specification of a further parameter $\eta$ that carries information about the sign of $\mbox{Re}e^{{\textrm{i}p/2\theta_{[j]}}}$.
Specifically $\eta$ indicates the fraction of phases with $\theta\in (-\pi,\pi]$ while $1-\eta$ indicates the fraction of phases with $\theta\in [\pi,3\pi)$.
When $\eta$ is known, we have that, assuming  $X_{p/2}$ real, the expression of $X_{p/2}$ is given for $p\in \{1,3,5\}$, by  
\bea
X_{1/2}&=&(2\eta-1)\int_{|\omega|\leq \Lambda} d\omega g(\omega) \sqrt{\frac{1}{2}\left[1+\sqrt{1-\left(\frac{\omega}{\Lambda}\right)^2}\right]}\nonumber \\
X_{3/2}&=&(2\eta-1)\int_{|\omega|\leq\Lambda} d\omega g(\omega) \left[\left( 2  \sqrt{1 - \left(\frac{\omega}{\Lambda}\right)^2} - 1\right) \sqrt{\frac{1 + \sqrt{1 - \left(\frac{\omega}{\Lambda}\right)^2}}{2}}\right]\nonumber  \\
X_{5/2}&=& (2\eta-1)\int_{|\omega|\leq \Lambda} d\omega g(\omega) \left[\left(3 - 4\left(\frac{\omega}{\Lambda}\right)^2 - 2\sqrt{1 - \left(\frac{\omega}{\Lambda}\right)^2} \right) \sqrt{\frac{1 + \sqrt{1 - \left(\frac{\omega}{\Lambda}\right)^2}}{2}} \right]
\eea
Note that for Gaussian as well as for the Lorentzian distribution of intrinsic frequencies, these integrals need to be evaluated numerically.

\section{Critical properties for $M=3$}

The \emph{saddle-node} bifurcation occurring at $\sigma_n=\sigma_n^{SN}$ can be found by imposing that for a non zero value of $\Lambda^\star$ we have
\begin{equation}
\Lambda^\star=f(\Lambda^\star;\{\sigma_n^{SN}\}),\qquad 1=f'(\Lambda^\star;\{\sigma_n^{SN}\}),\qquad f''(\Lambda^\star;\{\sigma_n^{SN}\})\neq 0 .
\label{eq:SN}
\end{equation}
By expanding around this solution for $|\delta\Lambda|=|\Lambda-\Lambda^\star|\ll 1$ and $\sigma_n=\sigma_n^{SN}+\delta\sigma_n$ with $|\delta\sigma_n|\ll 1$ one obtains
\begin{equation}
\delta\Lambda=f'(\Lambda^\star;\{\sigma_n^{SN}\})(\delta\Lambda)+\frac{1}{2!}f''(\Lambda^\star;\{\sigma_n^{SN}\})(\delta\Lambda)^2+\sum_n\left.\frac{\partial f(\Lambda;\{\sigma_k\})}{\partial\sigma_n}\right|_{\Lambda=\Lambda^\star,\sigma_k=\sigma_k^{SN}}\delta\sigma_n+\ldots
\end{equation}
Thus using the condition Eq.~(\ref{eq:SN}) one obtains
\begin{equation}
\frac{1}{2!}f''(\Lambda^\star;\{\sigma_n^{SN}\})(\delta\Lambda)^2=-\sum_n\left.\frac{\partial f(\Lambda;\{\sigma_k\})}{\partial\sigma_n}\right|_{\Lambda=\Lambda^\star,\sigma_k=\sigma_k^{SN}}\delta\sigma_n .
\end{equation}
Thus by approaching the saddle node from the synchronized state, the transition is hybrid with a square root critical behavior with $(\Lambda-\Lambda^\star)^2$ scaling linearly with $\delta\sigma_n$, i.e.\ $\Lambda-\Lambda^\star\propto|\delta\sigma_n|^{1/2}$.

The incoherent phase $\Lambda=0$ loses stability at a \emph{pitchfork} that occurs when $\sigma_n=\sigma_n^P$ with
\begin{equation}
f(0;\{\sigma_n^P\})=0,\qquad 1=f'(0;\{\sigma_n^P\}),
\label{eq:P}
\end{equation}
where the first condition is always satisfied since $X_{p/2}\to 0$ for $\Lambda\to 0$.
The Taylor expansion of $\Lambda$ around $\Lambda=0$, i.e.\ for $\Lambda=\delta\Lambda$ with $|\delta\Lambda|\ll 1$ and $\sigma_n=\sigma_n^P+\delta\sigma_n$, is in general given by
\begin{equation}
\delta\Lambda=f'(0;\{\sigma_n^P\})(\delta\Lambda)+\frac{1}{2!}f''(0;\{\sigma_n\})(\delta\Lambda)^2+\frac{1}{3!}f'''(0;\{\sigma_n^P\})(\delta\Lambda)^3+\ldots,
\end{equation}
with $f'(0;\{\sigma_n^P\})=1+\epsilon$ and $\epsilon=\sum_n\partial_{\sigma_n}f'(0;\{\sigma_k^P\})\,\delta\sigma_n$.
Dividing by $\delta\Lambda\neq 0$ this reduces to
\begin{equation}
0=\epsilon+\frac{1}{2!}f''(0;\{\sigma_n^P\})(\delta\Lambda)+\frac{1}{3!}f'''(0;\{\sigma_n^P\})(\delta\Lambda)^2+\ldots
\label{eq:exp}
\end{equation}
In the generic case, $f''(0;\{\sigma_n^P\})\neq 0$, we obtain $\delta\Lambda\simeq-2\epsilon/f''(0;\{\sigma_n^P\})$: for $f''(0;\{\sigma_n^P\})<0$ the system is in the supercritical region with $\delta\Lambda\propto\epsilon$, while for $f''(0;\{\sigma_n^P\})>0$ the system is in the subcritical region.

The {\em cusp} is found  for $\sigma_n=\sigma_n^C$ when we have
\begin{equation}
f(0;\{\sigma_n^C\})=0,\qquad 1=f'(0;\{\sigma_n^C\}),\qquad f''(0;\{\sigma_n^C\})=0,\qquad f'''(0;\{\sigma_n^C\})<0 .
\label{eq:cusp}
\end{equation}
At the cusp Eq.~(\ref{eq:exp}) reduces to $0=\epsilon+\frac{1}{3!}f'''(0;\{\sigma_n^C\})(\delta\Lambda)^2$, leading to the critical scaling
\begin{equation}
\delta\Lambda\simeq\left[-\frac{6\,\epsilon}{f'''(0;\{\sigma_n^C\})}\right]^{1/2}\propto\epsilon^{1/2}.
\end{equation}

However, when only integer order parameters enter $f$ (i.e.\ for $h_1$ and $h_2$ odd), $f$ is odd in $\Lambda$, since $X_k$ has parity $(-1)^k$ and the indices of every product in Eq.~(16) sum to one. Therefore $f''(0;\{\sigma_n^C\})=f''''(0;\{\sigma_n^C\})=0$ identically and Eq.~(\ref{eq:exp}) becomes
\begin{equation}
0=\epsilon+\frac{1}{3!}f'''(0;\{\sigma_n^C\})(\delta\Lambda)^2+\frac{1}{5!}f^{(5)}(0;\{\sigma_n^C\})(\delta\Lambda)^4+\ldots
\label{eq:expodd}
\end{equation}
Then for $f'''(0;\{\sigma_n^C\})<0$ the system is in the supercritical region with $\delta\Lambda\propto\epsilon^{1/2}$ (as in the standard Kuramoto model), while for $f'''(0;\{\sigma_n^C\})>0$ the system is in the subcritical region.
The cusp is found when we have
\begin{equation}
1=f'(0;\{\sigma_n^C\}),\qquad f'''(0;\{\sigma_n^C\})=0,\qquad f^{(5)}(0;\{\sigma_n^C\})<0,
\label{eq:cuspodd}
\end{equation}
and at the cusp Eq.~(\ref{eq:expodd}) leads to the critical scaling
\begin{equation}
\delta\Lambda\simeq\left[-\frac{120\,\epsilon}{f^{(5)}(0;\{\sigma_n^C\})}\right]^{1/4}\propto\epsilon^{1/4}.
\end{equation}
For $M=2$ and Lorentzian distribution of the intrinsic frequencies, Eq.~(\ref{eq:cuspodd}) gives the cusps $(\sigma_1,\sigma_2)=(-4\Delta,8\Delta)$ for $h=3$ and $(-2\Delta,32\Delta/5)$ for $h=5$ reported in Table~I.

Finally, when half-integer order parameters enter $f$ (i.e.\ $h_1$ or $h_2$ even), $f''(0;\{\sigma_n^C\})\propto(2\eta-1)^2$. Since $\eta=1/2$ in the incoherent state, on the synchronization path $f''(0;\{\sigma_n^C\})=0$ and the onset of synchronization has $\delta\Lambda\propto\epsilon^{1/2}$, while $\eta\neq 1/2$ can only be maintained on the desynchronization path for $\sigma_2>0$, where $f''(0;\{\sigma_n^C\})>0$ and the transition is subcritical.

\section{Critical behavior for the $M=2$ case}
\subsection{General expression of the Cup and Cap dynamics for $M=2$}
For  $M=2$, the Cup and Cap Synchronization  on a fully connected network is driven by the following dynamical equations 
\bea
\frac{d\theta_{[j]}}{dt}=\omega_{[j]}-\mbox{Im}\left[\Lambda e^{\textrm{i} \theta_{[j]}}\right].
\eea
with $\Lambda$ given by 
\bea
\Lambda = \sigma_1 \Theta^{(1)} + \frac{\sigma_2}{2^h}\Theta^{(2)}
\eea
with $\Theta^{(1)}=X_1$ and $\Theta^{(2)}$ depending on $h_1=h$ given by 
\bea
\Theta^{(2)}=\sum_{n=0}^{h_1} \binom{h_1}{n} X_{(2n-h_1+1)/2}X_{(-2n+h_1+1)/2}
\eea
It might be illustrative to express $\Theta^{(2)}$ for different values of $h$, i.e. $h\in \{0,1,2,3,4,5\}$:
\bea
\Theta^{(2)}=\left\{
\begin{array}{lll}
 {X}_{1/2}^2 &\mbox{for} &h=0 \\
2X_{1} &\mbox{for} &h=1 \\
2X_{-1/2}X_{3/2}+2X_{1/2}^2&\mbox{for} &h=2 \\
2X_{-1}X_{2}+6X_{1}&\mbox{for} &h=3 \\
2X_{-3/2}X_{5/2}+8X_{-1/2}X_{3/2}+6X_{1/2}^2&\mbox{for} &h=4  \\
2X_{-2}X_{3} + 10 X_{-1}X_{2} + 20X_{1} &\mbox{for} & h=5
\end{array}\right.
\eea
\subsection{Supplementary results for Gaussian intrinsic frequency distribution}
\begin{figure*}[ht!]
    \includegraphics[width=0.8\textwidth]{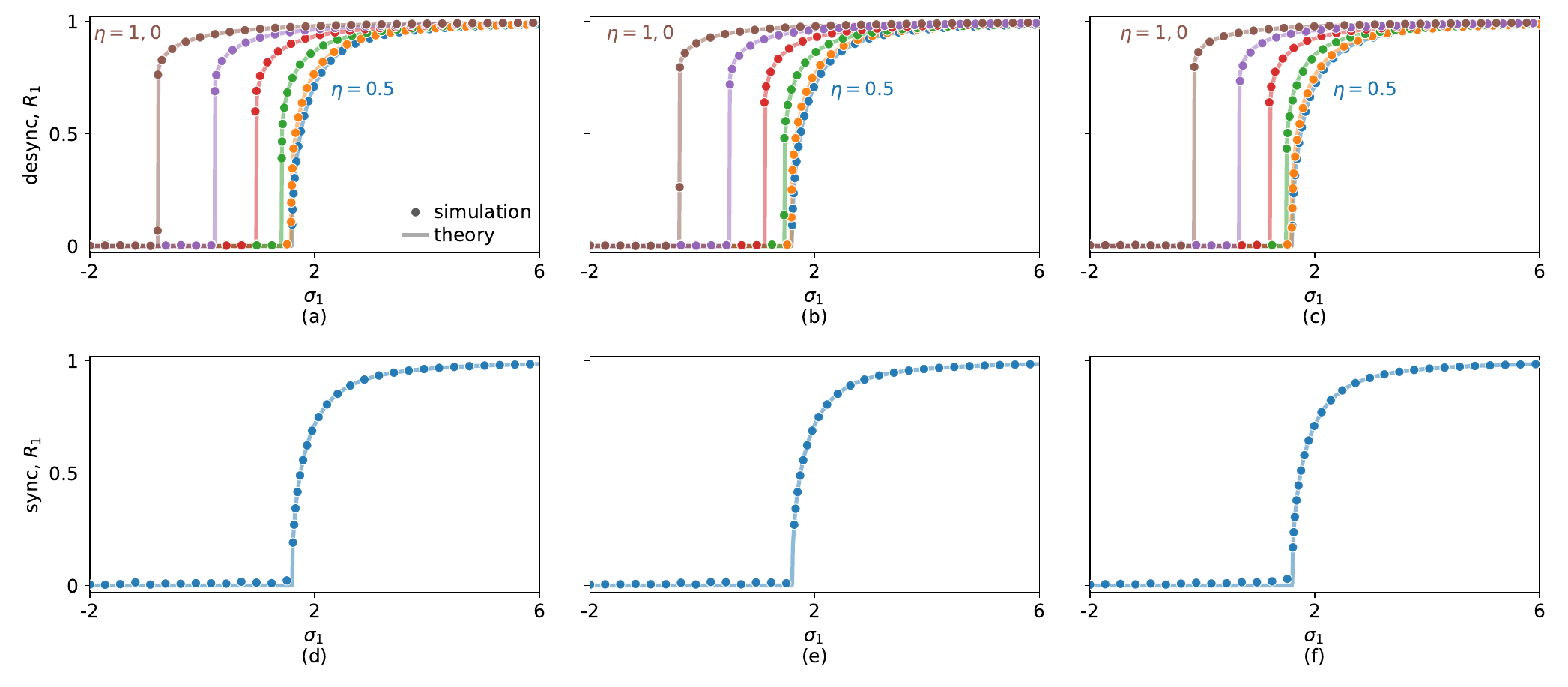}
    \caption{Synchronization and desynchronization on fully connected networks for even $h$. The order parameter $R_1$ as a function of $\sigma_1$ for different values of $h$ and of the bias $\eta,$ at $\sigma_2 = 3$ and Gaussian-distributed intrinsic frequencies. Panels (a) and (d), (b) and (e), and (c) and (f) correspond to $h = 0, 2, $ and $ 4$ respectively. Panels (a)-(c) show the desynchronization by decreasing $\sigma_1,$ for $\eta$ or $1-\eta$ equal to $1$ (brown), $0.9$ (purple), $0.8$ (red), $0.7$ (green), $0.6$ (orange) and $0.5$ (blue). Panels (d)-(f) show the synchronization by increasing $\sigma_1$ with $\eta = 0.5.$ The simulation results  on a fully connected network with $N=10^4$ nodes are shown in symbols (dots) while the theoretical results are indicated by lines.}
    \label{fig:even}
\end{figure*}

\begin{figure*}[!hb!]
    \includegraphics[width=0.8\columnwidth]{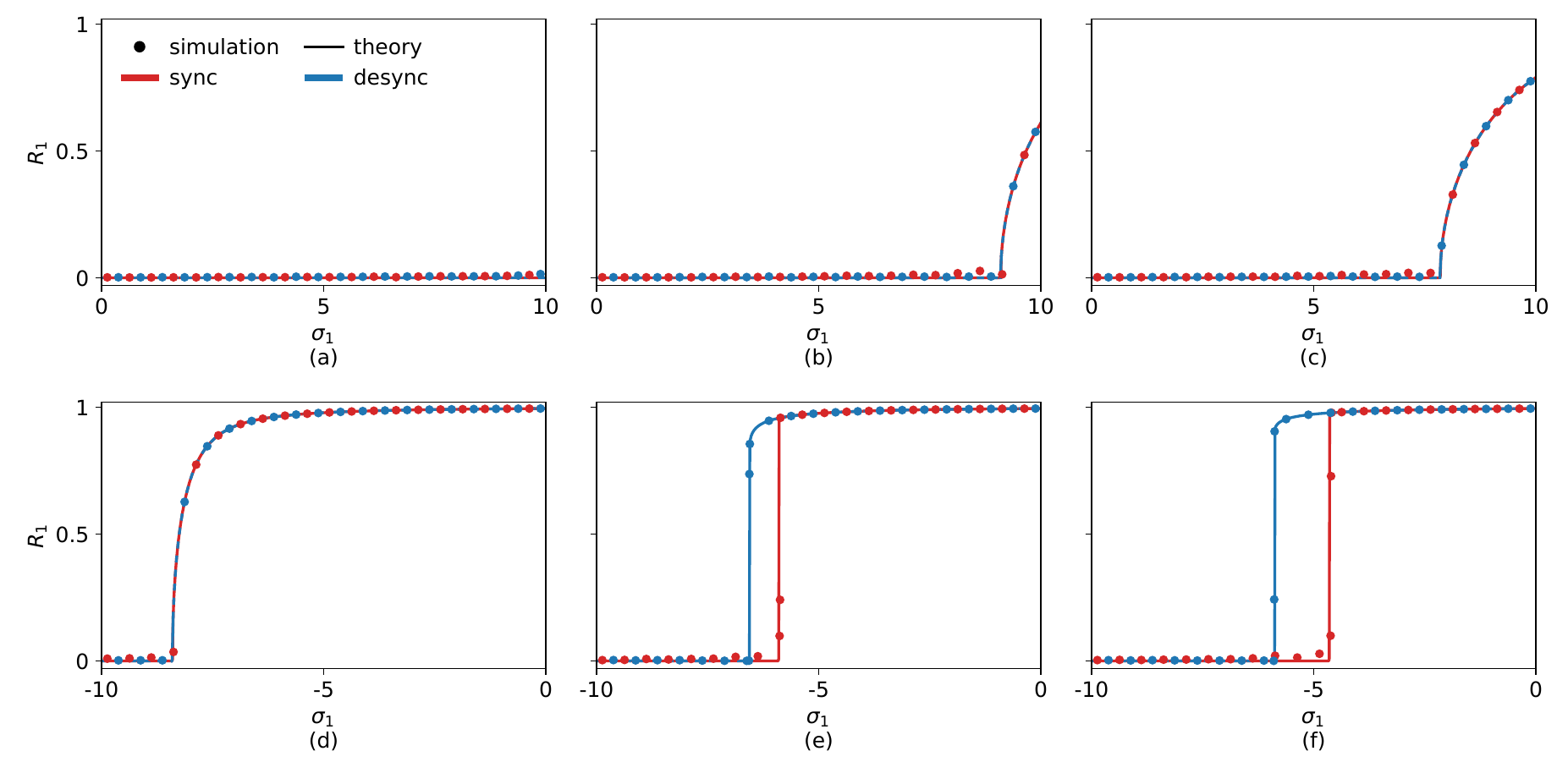}
    \caption{Synchronization and desynchronization on fully connected networks for odd $h$. The order parameter $R_1$ as a function of $\sigma_1$ for different values of $h$ and $\sigma_2$ with Gaussian-distributed intrinsic frequencies.  Panels (a) and (d), (b) and (e), and (c) and (f) correspond to $h = 1, 3, $ and $ 5$ respectively, with $\sigma_2 = 10$  in panels (a)-(c), and $\sigma_2 = -10$ in panels (d) - (f). The oscillators are first synchronized by increasing $\sigma_1$, then  desynchronized by decreasing $\sigma_1$. Hysteresis is observed in panels (e) and (f). The simulation results  on a fully connected network with $N=10^4$ nodes are shown in symbols (dots) while the theoretical results are indicated by lines.}
    \label{fig:odd}
\end{figure*}

The phase diagrams obtained for  standard normal distribution of intrinsic frequencies $\omega\sim\mathcal{N}(0,1)$ and $M=2$ are shown in the main body of the paper.
Here we  provide further numerical results of synchronization and desynchronization transitions obtained for the choice of $h_1=h$ even (see Figure \ref{fig:even}) and $h_1=h$ odd (see Figure \ref{fig:odd}). Notice that for $h$ even the initial condition is such that the phases for large value of $\sigma_1$ are initially drawn with probability $\eta$, uniformly at random with distribution $U[0,2\pi)$ and with probability $1-\eta$ uniformly at random with distribution $U[2\pi,4\pi)$.

\subsection{Supplementary results on  Lorentzian intrinsic frequency distribution case}

\begin{figure*}[!hb]
    \includegraphics[width=0.8\textwidth]{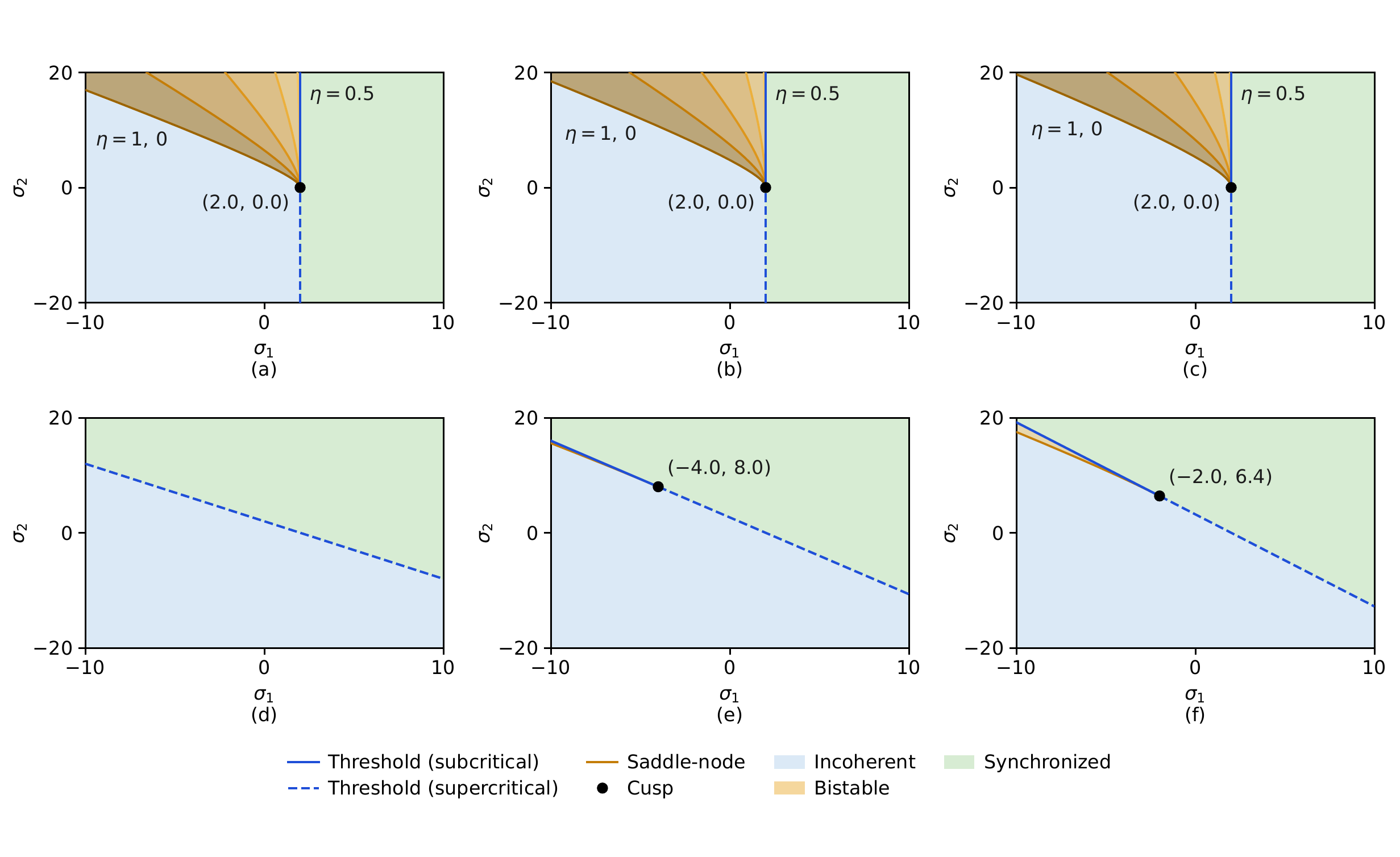}
   \caption{Phase diagrams of the cup and cap higher-order synchronization model on fully connected networks with Lorentzian distributed intrinsic frequencies, as a function of $h$ and coupling constants $\sigma_1$ and $\sigma_2$. Panels (a)-(c) correspond to $h = 0, 2, 4$. Brown lines, from dark to light, correspond to the saddle-node bifurcations with bias $\eta, (1-\eta) = 1, 0.9, 0.8, 0.7, 0.6, 0.5$. Panels (d)-(f) correspond to $h = 1, 3, 5$}.
    \label{fig:phasediagramL}
\end{figure*}

The phase diagrams obtained for $M=2$ and Lorentzian intrinsic frequency distribution $g(\omega)$ given by Eq.(\ref{Lorentzian}) are shown in Figure \ref{fig:phasediagramL}.
As is apparent from these phase diagrams, the same distinct critical behavior is observed also for Lorentzian intrinsic frequency distributions for choices of $h$ even and $h$ odd. For $h$ odd the pitchfork curve, the saddle-node curve and the cusp can be predicted analytically and are shown in Table \ref{tableL}. 

\begin{figure*}[!ht]
    \includegraphics[width=0.8\columnwidth]{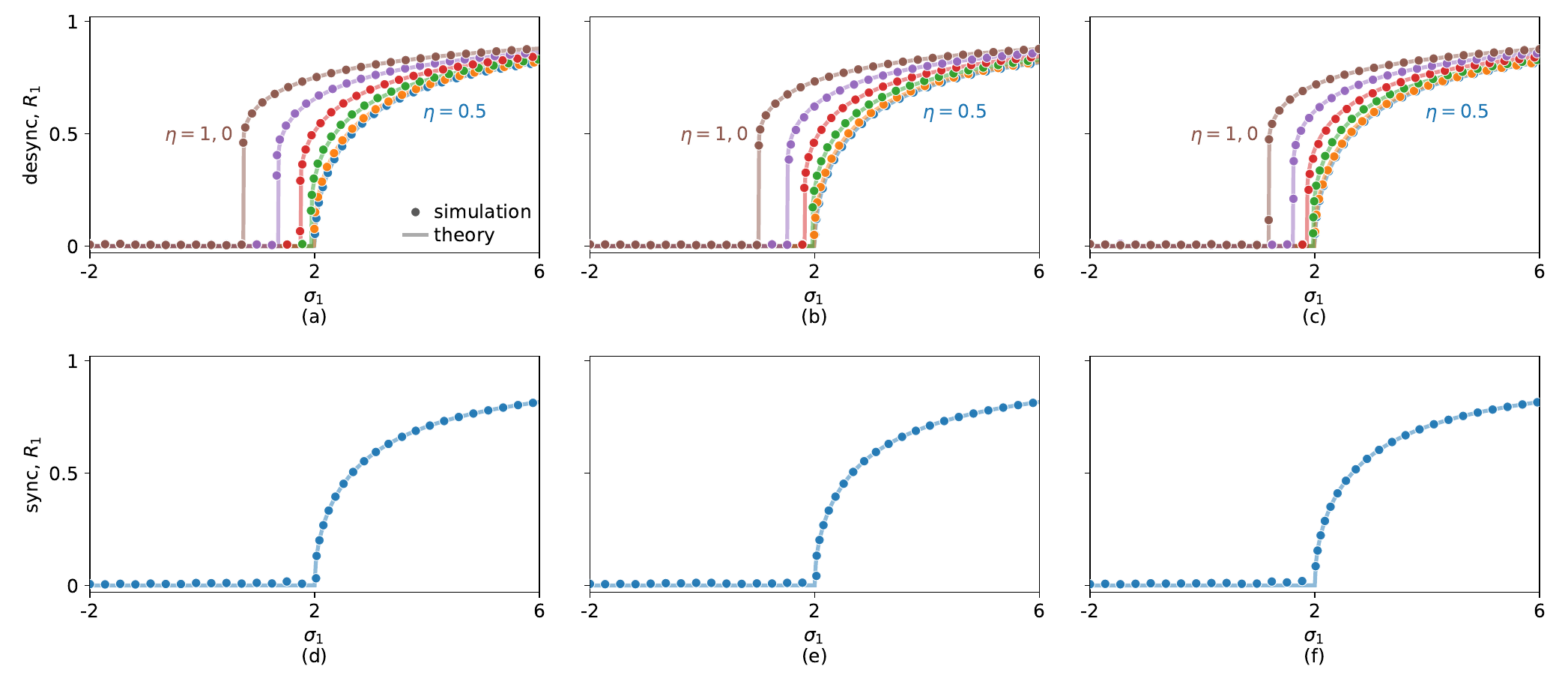}
    \caption{Synchronization and desynchronization on fully connected networks for even $h$. The order parameter $R_1$ as a function of $\sigma_1$ for different values of $h$ and of the bias $\eta,$ at $\sigma_2 = 3$ and Lorentzian-distributed intrinsic frequencies. Panels (a) and (d), (b) and (e), and (c) and (f) correspond to $h = 0, 2, $ and $ 4$ respectively. Panels (a)-(c) show the desynchronization by decreasing $\sigma_1,$ for $\eta$ or $1-\eta$ equal to $1$ (brown), $0.9$ (purple), $0.8$ (red), $0.7$ (green), $0.6$ (orange) and $0.5$ (blue). Panels (d)-(f) show the synchronization by increasing $\sigma_1$ with $\eta = 0.5.$ The simulation results  on a fully connected network with $N=10^4$ nodes are shown in symbols (dots) while the theoretical results are indicated by lines.}
    \label{fig:evenL}
\end{figure*}

\begin{figure*}[!hb!]
    \includegraphics[width=0.8\columnwidth]{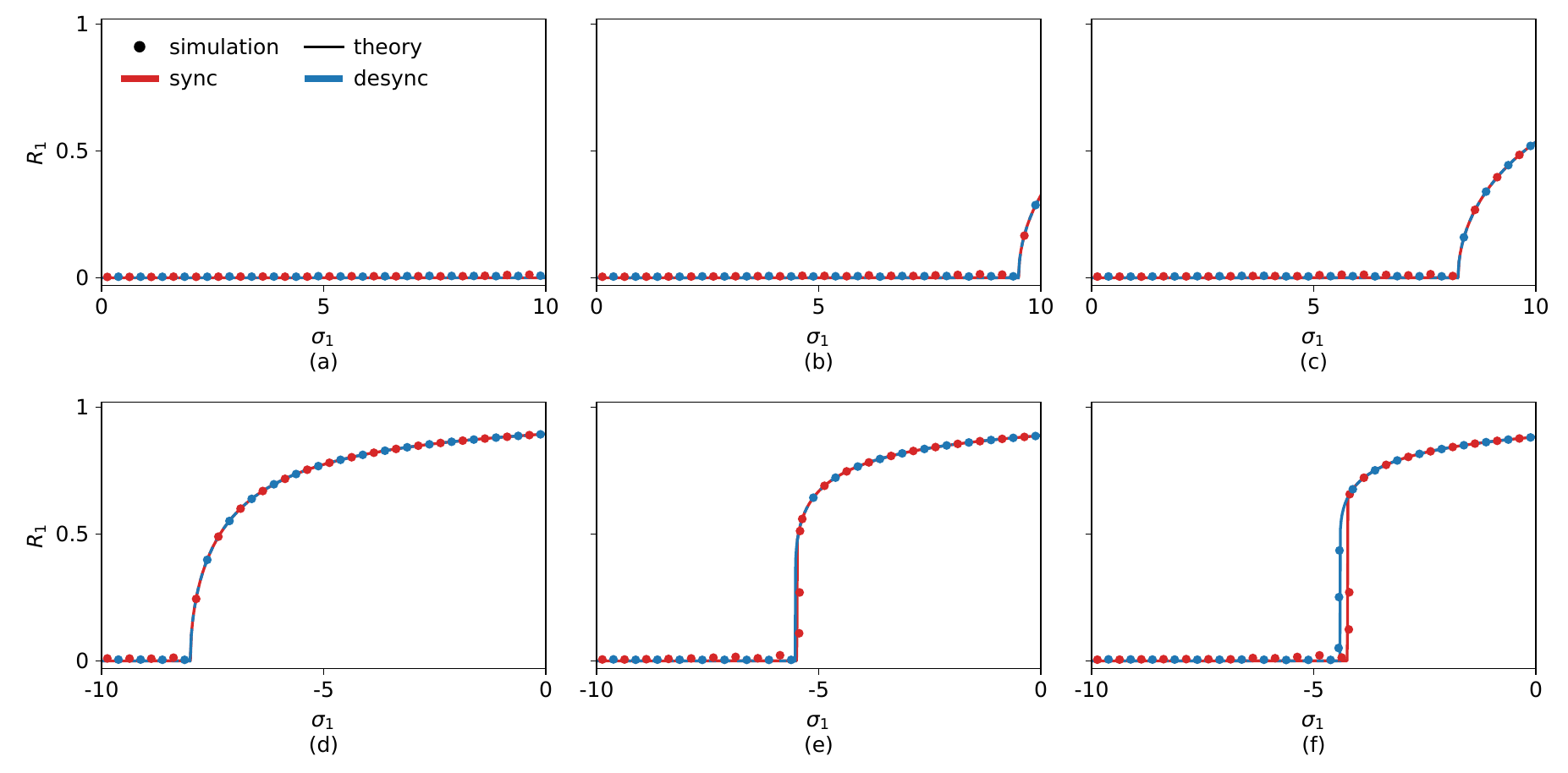}
    \caption{ Synchronization and desynchronization on fully connected networks for odd $h$. The order parameter $R_1$ as a function of $\sigma_1$ for different values of $h$ and $\sigma_2$ with Lorentzian-distributed intrinsic frequencies.  Panels (a) and (d), (b) and (e), and (c) and (f) correspond to $h = 1, 3, $ and $ 5$ respectively, with $\sigma_2 = 10$  in panels (a)-(c), and $\sigma_2 = -10$ in panels (d) - (f). The oscillators are first synchronized by increasing $\sigma_1$, then  desynchronized by decreasing $\sigma_1$. Hysteresis is observed in panels (e) and (f). The simulation results  on a fully connected network with $N=10^4$ nodes are shown in symbols (dots) while the theoretical results are indicated by lines.}
    \label{fig:oddL}
\end{figure*}

To confirm our theoretical results we also include here  numerical evidence of the good match between theory and simulations  for the  synchronization and the desynchronization transitions obtained for the choice of $h_1=h$ even (see Figure \ref{fig:evenL}) and $h_1=h$ odd (see Figure \ref{fig:oddL}). Notice that for $h$ even the initial condition is such that the phases for large value of $\sigma_1$ are initially drawn with probability $\eta$, uniformly at random with distribution $U[0,2\pi)$ and with probability $1-\eta$ uniformly at random with distribution $U[2\pi,4\pi)$.

\begin{table}[htbp]
\centering
\renewcommand{\arraystretch}{2.0}
\begin{tabular}{||c|c|c|c||}
\hline\hline
$h$ & Pitchfork / transcritical line & Saddle-node curve, $R=R_1$ & Cusp $(\sigma_1,\sigma_2)$\\
\hline
$1$ & $\sigma_1+\sigma_2=2\Delta$ & none (always supercritical) & none\\
\hline
$3$ & $\sigma_1+\dfrac34\sigma_2=2\Delta$ &
\begin{tabular}{@{}c@{}}$\sigma_1^{\rm SN}=-\dfrac{4\Delta(1+R^2)}{(1-R^2)^2}$\\ $\sigma_2^{\rm SN}=\dfrac{8\Delta}{(1-R^2)^2}$\end{tabular} & $(-4\Delta,\,8\Delta)$\\
\hline
$5$ & $\sigma_1+\dfrac58\sigma_2=2\Delta$ &
\begin{tabular}{@{}c@{}}$\sigma_1^{\rm SN}=-\dfrac{2\Delta(3R^4+8R^2+5)}{(1-R^2)^2(2R^2+5)}$\\ $\sigma_2^{\rm SN}=\dfrac{32\Delta}{(1-R^2)^2(2R^2+5)}$\end{tabular} & $\left(-2\Delta,\,\dfrac{32}{5}\Delta\right)$\\
\hline\hline
\end{tabular}
\caption{Table indicating the pitchfork/transcritical line, the saddle-node curve,  and the  cusp coordinates $(\sigma_1,\sigma_2)$ as a function of the order parameter $R=R_1$ for the Cup and Cap Synchronization with $M=2$, $h$ odd, with $h=1,3,5$,  and Lorentzian intrinsic frequency distribution.}
\label{tableL}
\end{table}

\end{document}